\documentclass[aps,prd,reprint,showpacs,twocolumn]{revtex4}
\usepackage{amsmath,amssymb,amsthm,graphicx}
\usepackage[mathscr]{eucal}
\usepackage[colorlinks,linkcolor=blue,
anchorcolor=blue,citecolor=green]{hyperref}
\usepackage[dvipsnames,usenames]{color}

\usepackage{epsf,epsfig,graphics}
\usepackage{verbatim,color,ulem}
\newcommand{\vx}{\ensuremath{\vec{x}}}
\newcommand{\vk}{\ensuremath{\vec{k}}}

\newcommand{\be}{\begin{equation}}
\newcommand{\ee}{\end{equation}}
\newcommand{\bea}{\begin{eqnarray}}
\newcommand{\eea}{\end{eqnarray}}

\begin{document}
\title{Quantum loop effects to the power spectrum of primordial perturbations during ultra slow-roll inflation}

\author{Wei-Can Syu}
\affiliation{Department of Physics, National Dong Hwa University, Hualien 97401, Taiwan, Republic of China}
\author{Da-Shin Lee}
\email{dslee@gms.ndhu.edu.tw}
\affiliation{Department of Physics, National Dong Hwa University, Hualien 97401, Taiwan, Republic of China}
\author{Kin-Wang Ng}
\email{nkw@phys.sinica.edu.tw}
\affiliation{Institute of Physics, Academia Sinica, Taipei 11529, Taiwan,  Republic of China}
\affiliation{Institute of Astronomy and Astrophysics,
Academia Sinica, Taipei 11529, Taiwan, Republic of China}
\date{\today}

\begin{abstract}
We examine the quantum loop effects on the single-field inflationary models in a spatially flat Friedmann-Robertson-Walker cosmological space-time with a general self-interacting scalar
field potential, which is  modeled  in terms of the Hubble flow parameters in the effective field theory approach. In particular, we focus on the scenarios in both slow-roll to ultra-slow-roll (SR-USR) and SR-USR-SR inflation, in which it is shown that density perturbations originating from  quantum vacuum  fluctuations can be enhanced at small scales, and then potentially collapse into primordial black holes. Here, 
 our estimates  indicate
significant one-loop corrections around the peak of the density power spectrum in both scenarios.
The induced large quantum loop effects
should be confirmed by a more formal quantum field theory and, if so, should be treated in a self-consistent manner
that will be discussed.
\end{abstract}

\maketitle

\section{Introduction}
Primordial black holes (PBHs) have recently received  renewed attention since  the discovery of the gravitational waves emitted by the merging of two 30 $M_{\odot}$  black holes, speculated to be PBHs  resulting in LIGO coalescences \cite{ABB}.
In addition, the focus has been on the possibility that if PBHs are abundant enough they could comprise a considerable fraction of
the dark matter and thus leave imprints throughout the history of the Universe \cite{CAR,MES,CAR1} (see Ref. \cite{SAS} for a review).

Large-scale structures of the Universe are seeded by quantum vacuum fluctuations during the very early evolution of the Universe and then stretched to cosmological scales by the rapidly exponential expansion of the inflation.  During such  a stage of primordial
acceleration, the curvature perturbation originating from  quantum vacuum  fluctuations may be enhanced at small scales with respect to the large-scale perturbations, which are ultimately responsible for the cosmic microwave background anisotropies. At cosmological horizon reentry, the small-scale
fluctuations in the overdense region might collapse into a PBH if they are large enough to overcome the pressure gradients.
Nevertheless, such an enhancement on small-scale fluctuations can occur either within single field of models of inflation or through some spectator field.  The models of extensive studies include critical
Higgs inflation \cite{EZQcrit}, double inflation \cite{KAN}, radiative plateau inflation \cite{BAL}, and some string realizations \cite{CIC,OZS,DAL}, to cite a few.  Generally speaking, for having sufficiently large fluctuations, the
inflationary dynamics has to deviate from slow-roll (SR) \cite{GER,hu}.  Ultra-slow-roll (USR) inflation has been proposed  as  a transient  phase of single-field inflation to generate large small-scale perturbations \cite{KIN,MAR,CHE,BYR}. The idea of USR is to consider a very flat potential of the inflaton field when its equation of motion, given by the Klein-Gordon equation, in a Friedmann-Robertson-Walker (FRW) cosmological space time, is dominated by the cosmological friction term rather than the slope of the potential in the SR inflation case. Although the slope of the potential is very small, we still have potential domination in  the Friedmann equations, so inflation continues.
If so, the time derivative of the inflaton field becomes exponentially small in time, thus
enhancing the produced curvature perturbations, which are inversely proportional to the inflaton velocity.
However, in this scenario for very small inflaton velocity, small quantum kicks due to quantum loop effects might become comparable to its mean value. A natural framework to involve quantum kicks is through the stochastic inflation \cite{STA}. Several works \cite{FIR,PAT,BIA,EZQ} have been devoted to the study of this quantum noise effects and have found a significant boost of PBH production, whereas the work in Ref.~\cite{CRU} has claimed that quantum diffusion effects in the USR inflation are insignificant by keeping the formalism in its regime of validity.

To see the quantum loop effects on the curvature perturbations, perhaps giving some implications to the above controversy, in this article, we will perform the loop quantum field theory calculations by mainly following the work of Ref.~\cite{boyan1} and  consider the loop effects  from the quantum fluctuations of the inflaton field itself. In particular, the quantum field fluctuations of $ \langle \varphi^2 \rangle$ not only have the usual ultraviolet divergence in its one-loop momentum integral that can be removed by a proper procedure of regularization/renormalization by defining its renormalized counterpart, but also in the case of minimally coupled massless inflaton fluctuations in de Sitter space-time, they suffer from the infrared divergence~\cite{fordbunch}. In Ref.~\cite{boyan1}, its infrared enhancement that gives sizable effects to modify the slow roll parameters was studied. Here, we will extend the study to the USR inflation.

Our paper is organized as follows. In next section,  we  introduce single-field inflationary models in a metric of the perturbed spatially flat FRW cosmological space-time in
the
Arnowitt-Deser-Misner (ADM) form and then model the general self-interacting scalar
field  potential $V(\phi)$  in terms of the Hubble flow parameters in the effective field theory approach. We then separate  the classical homogeneous background field ($\Phi_0$)  from the
quantum field fluctuations ($\varphi$).
In a spatially flat gauge,
the background field  in the FRW metric with the one-loop corrections is derived, whereas
the scale factor follows the modified Friedmann equations also including the one-loop contributions.  The equation of motion for mode functions of the quantum field fluctuations is derived, and the solution of the Hankel function of order $\nu$ is found.
Later,  the Bunch-Davies vacuum state is chosen  to compute the one-loop effects of $\langle \varphi^2 \rangle$
that will backreact  the dynamics of the background field.
 In Sec.~\ref{sec3}, we introduce the power spectrum of primordial perturbations  described by the density perturbations in a spatially flat gauge.
 We obtain the one-loop expressions of the density perturbations as well as the energy density and pressure  of the inflaton field and consider their loop corrections given by  the potentially dominated $\langle \varphi^2 \rangle$ term due to its infrared enhancement as the order of the Hankel function $\nu \rightarrow 3/2$ during the inflationary epoch.
 In Sec.~\ref{sec4}, we first adopt the SR step model proposed in Ref. \cite{hu} to numerically study the SR to USR inflation and then modify the model to consider the SR-USR-SR inflation.
 We show that both scenarios can produce large density perturbations. We then study the effects from the one-loop contributions.
Concluding remarks and discussions are in Sec.~\ref{sec5}.

\section{Effective field theory  and Hubble flow parameters }\label{sec2}
The single-field inflationary model
that we would like to explore is
described by a
general self-interacting scalar field theory in a curved space-time. The corresponding Einstein-Hilbert action with a minimally coupled  scalar field is given by
\bea\label{action}
S&=& S_g+S_{\phi}\nonumber\\
&=&\frac{1}{2} \int d^4 x \sqrt{-g} R
\nonumber\\
&&+\int d^4 x \sqrt{-g} \bigg[ - \frac{1}{2} \;
\partial_{\mu} \phi \, \partial^{\mu}\phi  -V(\phi) \bigg], \;
\eea
where $M_{Pl}^{-2}= 8 \pi G_N$ has been set to $M_{Pl}^{-2}=1$, which will be recovered for clarity.
The {generic} potential $V(\phi)$ will be parametrized later in terms of the Hubble flow parameters in an effective field theory approach
by following the work of Ref.~\cite{boyan1} and generalizing it to the USR inflation relevant to PBHs production.
The  metric for convenience is chosen in the ADM form in the $3+1$ decomposition as
\bea\label{metric}
ds^2&=&  g_{\mu\nu} dx^{\mu} dx^{\nu}\nonumber\\
&=&-{\mathcal N}^2 dt^2 + h_{ij} ( dx^i+ {\cal N}^i dt) ( dx^j+ {\cal N}^j dt)\, .
\eea
 ${\cal N}$ is the lapse function, and ${\cal N}^i$ is the shift vector, which are not dynamical but are Lagrangian multipliers.
The action  (\ref{action}) while substituting the metric (\ref{metric}) and  dropping out the boundary terms then becomes
\bea \label{action_adm}
S&=& \frac{1}{2} \int dt d^3 x \, \sqrt{h} \, \Big[ {\cal N}( R^{(3)} - 2V- h^{ij} \partial_i\phi \, \partial_j \phi)\nonumber \\
&&+\, {\cal N}^{-1} \big( E_{ij} E^{ij} -E^2 +
(\dot \phi- {\cal N}^i \partial_i \phi)^2 \big) \Big] \,,
\eea
where the extrinsic curvature can be obtained as $K_{ij}=E_{ij}/{\cal N}$ with $E_{ij}=({\dot h}_{ij}-\nabla_i {\cal N}_j-\nabla_j {\cal N}_i)/2$.
The scalar $E$ is the trace of $E=E^i_i$, while raising or lowering the spatial indices is via the metric $h_{ij}$. The overdot means the time derivative of the cosmic time $t$, and $\nabla_i$ is the covariant derivative also with respect to the metric $h_{ij}$. $R^{(3)}$ is the three-dimensional Ricci scalar  giving
 the intrinsic curvature.
 The equation of motion for ${\cal N}$ and ${\cal N}^i$ are the Hamiltonian and momentum constraints given, respectively, by
 \bea
R^{(3)}-2V -\frac{1}{{\cal N}^2} ( E_{ij} E^{ij} -E^2 )-\frac{1}{{\cal N}^2} (\dot \phi- {\cal N}^i \partial_i \phi )^2&&\nonumber\\
\hspace{5cm}-h^{ij} \, \partial_i \phi  \, \partial_j \phi =0,&& \nonumber\\
\nabla_i \bigg[ \frac{1}{{\cal N}} (E^i_j -\delta^i_j E) \bigg]-\frac{1}{{\cal N}} (\dot \phi-N^i \partial_i \phi) \partial_j \phi =0 ,&&
 \eea
 through  which we can solve for ${\cal N}$ and ${\cal N}_i$ by choosing a particular gauge and plug their solutions back to the action.
Moreover, we consider that the inflaton field is obtained from the homogeneous expectation value of the quantum scalar
field
\be\label{exp}
\Phi_0(t)=\langle \phi(\vx,t) \rangle
\ee
in a spatially flat
FRW cosmological space-time with  ${\cal N}=1$, ${\cal N}_i=0$, and $h_{ij}=a^2 (t) \delta_{ij}$, where $a$ is a scale factor. The expectation value is given by the nonequilibrium quantum state that later will
be specified to be the Bunch-Davies state often studied in the literature.
As such, the above momentum constraint is automatically satisfied.
Nevertheless, the Hamiltonian constraint gives the Friedmann equation
\bea
H_0^2 = \frac{\rho_{\Phi_0}}{3 \, M^2_{Pl}}\;  \label{hub0}
\;
\eea
with the Hubble parameter $H_0=\dot a/a$.  The corresponding energy density and pressure for   the spatially homogeneous part of the inflaton field are expressed as
\bea \label{rho_p}
&& \rho_{\Phi_0}= \frac{1}{2}(\dot{\Phi}_0)^2+
V(\Phi_0) \;,\nonumber\\
&& p_{\Phi_0}= \frac{1}{2}(\dot{\Phi}_0)^2-
V(\Phi_0)\;.
\eea
Apart from the dynamics of the scalar factor $a$, in the spatially flat FRW cosmology, the classical equation of motion for $\Phi_0$ can be derived from the action as
\be \label{phi0}
\ddot{\Phi}_0+3\,H_0\,\dot{\Phi}_0+V'(\Phi_0) =0 \, .
\ee
Here, we consider the perturbations about the above background fields of $\Phi_0$ and $H_0=\dot a/a$. It is found more convenient to work in the spatially flat gauge by writing ${\cal N}=1+ \delta {\cal N}$ and
\be\label{tad}
\phi(\vec{x},t)= \Phi_0(t)+\varphi(\vec{x},t)\;,
\ee
\noindent with
\be\label{exp}
\langle
\varphi(\vx,t)\rangle =0 \;,
\ee
where $\delta  {\cal N},\, {\cal N}_i$, and $\varphi$ are  perturbed fields.

According to Ref.~\cite{mal},
in order to expand the action (\ref{action_adm}) to quadratic and cubic order in the perturbed field $\varphi$, we only need to plug in the solution for the first-order perturbation in ${\cal N}$ and ${\cal N}_i$ and do the expansion
since the higher-order terms in ${\cal N}$ and ${\cal N}_i$  will be multiplying the Hamiltonian constraint or the variation of the Lagrangian with respect to ${\cal N}$ and ${\cal N}_i$, respectively,  to zeroth order, which all vanish.
We then expand out the Lagrangian density $S$~(\ref{action_adm}) up to the linear order in ${\cal N}$ and ${\cal N}_i$ as well as the cubic order in $\varphi$  with necessary integration by parts as \cite{mal,clu,kaz}
\bea\label{Split}
S = \int d^4 x \, a^3(t)\,
\mathcal{L}[\Phi_0(t),\varphi(\vx,t), \delta {\cal N}, {\cal N}_i ]\;,
\eea
\noindent with
\bea\label{lagra}
\mathcal{L}  &=&
\frac{1}{2} \; {\dot{\Phi}^2_0}-V(\Phi_0) \; \nonumber\\
&&- \varphi\;
\left[\ddot{\Phi}_0+3 \, H \,\dot{\Phi}_0+V^{'}(\Phi_0)\right]-\delta {\cal N} \, V (\Phi_0) \nonumber\\
&& + \frac{1}{2} {\dot{\varphi}^2}-\frac{\partial_i \varphi \, \partial_i \varphi}{2 \, a^2} -\frac{1}{2}\;
V^{''}(\Phi_0)\; \varphi^2 - V'(\Phi_0) \, \delta {\cal N} \, \varphi\nonumber\\
&&-\dot \Phi_0 \, \delta {\cal N} \, \dot \varphi -\frac{1}{6} V'''(\Phi_0) \varphi^3-\frac{1}{2} V''(\Phi_0) \, \varphi^2 \, \delta {\cal N}  \, \nonumber\\
&& -\frac{1}{2} \dot{\varphi}^2 \delta {\cal N} -\dot \varphi {\cal N}^i \partial_i \varphi-\frac{1}{2 a^2} \partial_i \varphi\,  \partial_i \varphi \, \delta {\cal N}\nonumber\\
&& + \, \text{higher-order terms} \,.
\eea
The first-order solutions are found as \cite{clu,kaz}
\bea \label{N}
&&\delta {\cal N}=\frac{1}{H_0 M^2_{Pl}} \bigg( \frac{\, \dot \Phi_0} {2} \varphi \, \bigg) \, ,\nonumber\\
&& {\cal N}_i=\partial_i \chi \,, \quad \partial^2 \chi=\frac{\dot{\Phi}_0^2}{ 2 \,  H_0^2 M_{Pl}^2} \frac{d}{dt} \bigg(-\frac{H_0}{\dot{\Phi}_0} \varphi \bigg) \,,
\eea
where  the Planck mass $M_{Pl}$ is put back to the expressions and is the largest energy scale in the framework of semiclassical gravity.
Notice that $\delta {\cal N}$ and ${\cal N}_i$ are all suppressed  not only by $1/M^2_{Pl}$ but also due to the smallness of $\dot\Phi_0$ during both SR and USR inflations, and they are
\be
\delta {\cal N} \propto \sqrt{\epsilon_1} \, , \quad \quad \quad {\cal N}_i \propto \sqrt{\epsilon_1} \,
\ee
with the small value of $\epsilon_1 =\frac{\dot \Phi_0^2} {2 M^2_{Pl} H_0^2} $ to be defined together with other Hubble flow parameters later. Henceforth, $\delta {\cal N}$ and ${\cal N}_i$ can be ignored  as compared with  loop effects from quantum field fluctuations of the scalar field of the order of $H_0/(M_{Pl}\sqrt{\epsilon_1})$  during the USR inflation to be seen later.

{To explore the effects of the quantum fluctuations of the scalar field,
the tadpole
method (see  Ref.~\cite{boyan3} and references
therein) is implemented to derive the equation of motion with the one-loop corrections for the homogeneous
expectation value of the inflaton field from the action (\ref{lagra})  by requiring the condition $\langle
\varphi(\vx,t)\rangle =0 $,
\be \label{phi_loop}
\ddot{\Phi}_0+3\,H \,\dot{\Phi}_0+V'(\Phi_0)+\frac{1}{2} V'''(\Phi_0) \, \langle \varphi^2 (\vec x,t) \rangle =0 \, ,
\ee
where $H$ obtained from the Hamiltonian constraint   with ${\cal N}=1$, ${\cal N}_i=0$, and $h_{ij}= a^2 \delta_{ij}$ again becomes
\bea \label{frieman_q}
H^2 &=&\frac{1}{3 M^2_{Pl}} \bigg\langle \frac{1}{2} {\dot{\phi}^2} +\frac{\partial_i \phi \partial_i \phi}{2 a^2} +V(\phi) \bigg\rangle \nonumber\\
&=&\frac{1}{3 M^2_{Pl}} \bigg\langle \frac{1}{2} {{\dot\Phi_0}}^2  +V(\Phi_0) \bigg\rangle\nonumber\\
&&+\frac{1}{3 M^2_{Pl}} \bigg\langle \frac{1}{2} {\dot{\varphi}}^2 +\frac{\partial_i \varphi \partial_i \varphi}{2 a^2} +\frac{1}{2} V''(\Phi_0) \varphi^2+ \cdot \cdot\cdot \bigg\rangle \nonumber\\
&=& H_0^2 +\delta H^2,
\eea
with the one-loop corrections to the Hubble parameter  $\delta H^2$ given by
\be \label{deltah}
\delta H^2=\frac{1}{3 M^2_{Pl}} \bigg\langle \frac{1}{2} {\dot{\varphi}}^2 +\frac{\partial_i \varphi \partial_i \varphi}{2 a^2} +\frac{1}{2} V''(\Phi_0) \varphi^2 \bigg\rangle \, .
\ee
Their quantum corrections will be estimated later after the proper renormalization to remove the UV divergence is done.}

Next, all the evolution of the Hubble parameter $H_0$ and the field $\Phi_0$ can be described  by the Hubble flow parameters defined as
\be
\epsilon_1=-\frac{\partial}{\partial N} \ln H_0  \, , \quad\quad \epsilon_{n+1}=\frac{\partial}{\partial N} \ln \epsilon_n \, , \label{HubFlow}
\ee
where $N=\ln a=\int \, H_0 \, dt$ is the number of $e$-folds.
Here, we will adopt the effective field theory approach  by directly modeling the effective potential $V(\phi)$ with
the Hubble flow parameters \cite{hu}.
The $\epsilon_n$ of relevance in this work
involves $\epsilon_1$ up to $\epsilon_4$, with their respective expression obtained as
\bea
\epsilon_1 &=&\frac{\dot \Phi_0^2} {2 M^2_{Pl} H_0^2}  \, ,\nonumber\\
\epsilon_2 &=&-6 \,\bigg( 1-\frac{\epsilon_1}{3}+\frac{V'(\Phi_0)}{3 H_0 \dot \Phi_0} \bigg) \, , \nonumber\\
\epsilon_3 &=&\frac{1}{\epsilon_2} \bigg( 5 \, \epsilon_1 \epsilon_2- 4 \,\epsilon^2_2 - 3\, \epsilon_2 + 12\, \epsilon_1 -\frac{\epsilon_2^2}{2} - \frac{2 \, V''(\Phi_0)}{H_0^2} \bigg), \nonumber\\
\epsilon_4 &=& -\epsilon_3 +\frac{1}{\epsilon_2\epsilon_3} \bigg( 6\, \epsilon_1 \epsilon_2^2 + 7 \, \epsilon_1 \epsilon_2 \epsilon_3 -18\, \epsilon_1^2 \epsilon_2-3 \,\epsilon_2 \epsilon_3 \nonumber\\
&& \quad+18 \,\epsilon_1 \epsilon_2-\epsilon_2^2 \epsilon_3 + 8 \,\epsilon_1^3 -24 \, \epsilon_1^2- \frac{2 \, V'''(\Phi_0) \dot\Phi_0}{H_0^3} \bigg) \, .\nonumber\\
\label{eps_V}
\eea
Given the Hubble flow parameters for undergoing the USR inflation to be shown in the figures later, we can choose $\vert \epsilon_2 \vert \sim {\mathcal{O}(1)}$ and $\vert \epsilon_2 \epsilon_3 \vert < 1$, but $|\epsilon_2\epsilon_3| > \epsilon_1$ for extremely small $\epsilon_1$, say,  $\epsilon_1 \sim 10^{-9}$, to be seen later.
Additionally,  all other $\epsilon$'s can be ignored.
{Then, the general effective potential $V(\Phi_0)$ and $\dot\Phi_0$ can be approximately reconstructed as
\bea \label{V_app}
\dot \Phi_0 &=& \sqrt{2 \epsilon_1} M_{pl} H_0 \, , \nonumber\\
V'(\Phi_0) & \simeq & - 3 \sqrt{2} M_{Pl} H_0^2 \sqrt{\epsilon_1} \bigg(1+ \frac{\epsilon_2}{6}\bigg)\, , \nonumber\\
V''(\Phi_0) &\simeq & - \frac{H_0^2}{2} \bigg(\epsilon_2 \epsilon_3+\frac{\epsilon_2^2}{2}+ 3 \epsilon_2 \bigg) \, , \nonumber\\
V'''(\Phi_0) & \simeq &- \frac{H^2_0}{\sqrt{ 2 \epsilon_1} M_{Pl}} \bigg( \frac{3}{2} \epsilon_2 \epsilon_3+ \frac{1}{2} \epsilon_2^2 \epsilon_3 \bigg) \,.
\eea
{We retain the terms up to the order $\epsilon_2 \epsilon_3$ in the  expressions of $V''(\Phi_0)$ and $V'''(\Phi_0)$ and the leading-order terms of $V'(\Phi_0)$.}
$V'(\Phi_0)$ is small as compared with $\dot \Phi_0 H_0$ as long as $1+\frac{\epsilon_2}{6}  <1$, whereas $\epsilon_2 \rightarrow -6$ is an extreme case.
The  approximate form of   $V'''(\Phi_0)$ surely results in the one-loop corrections
 in Eq.~(\ref{phi_loop}) of order  $\mathcal{O} (H^2_0/(M^2_{Pl}{\epsilon_1}))$. However,  $\delta H^2/H_0^2$ in Eq.~(\ref{deltah})  is of order $\mathcal{O} (H^2_0/M^2_{Pl})$ instead to be also ignored as compared with the loop corrections given by $V'''(\Phi_0)$ for extremely small $\epsilon_1$.}
 It is  anticipated that the effective field  theory approach relies on the separation between
the energy scale of inflation determined
by the Hubble parameter
 and the cutoff scale of  the Planck scale.
 The  expected dimensionless ratio of the effective field theory  approximation is the ratio $H_0/M_{Pl}$, which has to be small for safely ignoring the quantum gravity effects. Phenomenologically, the smallest $H_0/M_{Pl}$ gives a consistent constraint on the amplitudes of
tensor and scalar perturbations inferred from the Planck data~\cite{PLA}, thus leading to strong
observational support to the validity of an effective field theory for
inflation well below the Planck scale.
Now, we find it more convenient to
work in conformal time with the metric
background as
$$
ds^2= dt^2-a^2(t) \,d{\vec x}^2 = C^2(\eta) \left[ d \eta^2 -  d{\vec x}^2
\right] \; ,
$$
{where $ \eta $ is the  conformal time and the scale factor in a quasi-de Sitter space-time is { $C(\eta) \equiv a(t(\eta)) =-\frac{1}{H_0 \eta (1-\epsilon_1)}\simeq
-\frac{1}{H_0 \eta } (1+\epsilon_1)$ during the inflation for small $\epsilon_1$.}}
The conformally rescaled field is defined as
\be\label{rescale}
\varphi(\vx,t) =\frac{\chi(\vx,\eta)}{C(\eta)}\, ,
\ee
\noindent $ C(\eta) $ being the scale factor in conformal time.
 The spatial Fourier transform of the free-field Heisenberg
operator $\chi(\vx,\eta)$ obeys the equation, which can be read off from the quadratic terms in $\varphi$ in the Lagrangian density (\ref{lagra}) as
{
\bea\label{heiseqn}
 \chi^{''}_{\vk}(\eta)&+& \left[k^2 + [V''(\Phi_0)+ 2 V'(\Phi_0){\delta {\cal N}}/{\varphi}] 
 \;  C^2(\eta) \right.
 \nonumber\\
 &&\quad\quad \quad\quad \quad \left. -
\frac{C^{''}(\eta)}{C(\eta)} \right]\chi_{\vk}(\eta)=0\,.
\eea}
Here the prime means the derivative with respect to the conformal time $\eta$.
{Using the Hubble flow parameters  to express $V''(\Phi_0)$, $V'(\Phi_0)$, and $\delta {\cal N}$  in Eq.~(\ref{V_app}),}
{the mode equation becomes
\be\label{heiseqn2}
\chi^{''}_{\vk}(\eta)+ \left[k^2
-\frac{\nu^2-\frac{1}{4}}{\eta^2} \right]\chi_{\vk}(\eta)=0 \,,\ee
\noindent where the index $\nu$ can be approximated by
{
\bea\label{nu}
\nu^2 &\simeq &
\frac{9}{4}  +\bigg( \frac{3}{2}\, \epsilon_2 +\frac{1}{4} \, \epsilon_2^2\bigg) (1+2 \epsilon_1) \nonumber\\
&&\quad +\frac{1}{2} \,\epsilon_2 \epsilon_3+ 3 \, \epsilon_1+ \epsilon_1 \epsilon_2+6 \epsilon_1 \bigg(1+\frac{\epsilon_2}{6} \bigg) \,
\eea}
while retaining all terms in $\epsilon_2$ and the linear terms in $\epsilon_3$.
 The terms of $\epsilon_1$ dependence give very small contributions in the USR inflationary epoch that can certainly be ignored, but, in our work,
 we will explore the SR-USR and SR-USR-SR inflation and find that the evolution of quantum corrections is  sensitive to $\nu$. {However, during SR, the parameters we choose are
 $\epsilon_1 > \vert \epsilon_2\vert $ where $\epsilon_1, \vert \epsilon_2\vert \ll1$.  Thus, to accommodate the epoch of SR inflation in the models below, we also include the  linear $\epsilon_1$ terms in the expression of $\nu$ above.}  However, it can be checked that keeping the linear terms in $\epsilon_1$ in the approximate forms of $V''(\Phi_0)$ and $V'''(\Phi_0)$ in Eq.~(\ref{V_app}) will not have a sizable change in the numerical studies we perform later.
The scale-invariant case $ \nu = \frac{3}{2} $ corresponds to
massless inflaton fluctuations in the de Sitter background.
We then introduce the
quantity
 \be\label{delta} \Delta= \frac{3}{2}-\nu \, \ee
that measures the departure from scale invariance.
 The free
Heisenberg field operators $\chi_{\vk}(\eta)$ can be written in terms
of annihilation and creation operators   as
\be\label{ope}
\chi_{\vk}(\eta) = a_{\vk} \; S_{\nu}(k,\eta)+
a^{\dagger}_{-\vk} \; S^{*}_{\nu}(k,\eta)\,,
\ee
where the mode functions $S_{\nu}(k,\eta)$ are solutions of
Eq.~(\ref{heiseqn2}). The vacuum state from which  to build up the Fock space by acting the creation operator on it is
the the Bunch-Davies vacuum defined as
\be
a_{\vk} |0\rangle_{BD} =0 \, .
\ee
Then, these
mode functions are given by
\be\label{BDS} S_{\nu}(k,\eta) = \frac{1}{2}
\; \sqrt{-\pi\eta} \; e^{i\frac{\pi}{2}(\nu+\frac{1}{2})} \;
H^{(1)}_\nu(-k\eta)\, .
\ee
For large momenta $|k\eta| \gg 1$,
the mode functions behave the same as free-field modes in the Minkowski
space-time, i.e.,
\be \label{S_large_k}
S_{\nu}(k,\eta) = \frac{1}{\sqrt{2k}} \; e^{-ik\eta}\quad{\rm for}\quad|k\eta| \gg1\,.
\ee
In fact, in the theory of quantum fields in curved space,
there is no unique choice of a vacuum state.
 In this article, we focus on the standard
choice often adopted in the literature,  which allows us to include the
quantum corrections into the standard results in the literature. A study
of quantum loop corrections with different initial states is an important
aspect that deserves further study.
The index $\nu$ in the mode functions (\ref{BDS}) depends on the
expectation value of the scalar field,  via the Hubble flow parameters;
hence, it slowly varies in time. Therefore, it is  consistent to treat
this time  dependence of  $\nu$ as an {adiabatic approximation}.
This is well known and standard in the SR or USR
expansion.

Considering the Bunch-Davies vacuum state, the quantum correction $\langle \varphi^2 \rangle$  is given by
\bea\label{phi2}
\langle \varphi^2 (\vec x,t) \rangle && = \int \frac{d^3 k}{(2\pi)^3} \, \frac{\vert S_{\nu} (k,\eta) \vert^2}{C^2(\eta)}   \nonumber\\
&&= \frac{H^2}{8 \, \pi} \; \int^{\Lambda}_0 \frac{dk }{k} \, (-k\eta)^3 \;
\left| H^{(1)}_\nu(- k \eta)\right|^2 \, \nonumber\\
&&=\frac{H^2}{8 \, \pi}
\int^{\Lambda_p}_0 \frac{dz}{z} \; z^3\,
\left|H^{(1)}_\nu(z)\right|^2\,.
\eea
With the large $k$ behavior of the Hankel function in Eq.~(\ref{S_large_k}), the quantum correction $\langle \varphi^2 \rangle$  has UV divergence, which will be discussed in Appendix A, and thus can be dealt with by the proper regularization/renormalization procedure in Ref.~\cite{boyan1}.
Note that in the case of $\nu=3/2$ the
integrand in Eq.~(\ref{phi2}) becomes
\be\label{integ}
z^3 \,
\left|H^{(1)}_{3/2}(z)\right|^2 =
\frac{2}{\pi}\left[1+z^2\right]\,.
\ee
It is known that the integral of $\langle \varphi^2 \rangle$  has an additional
{infrared} logarithmic divergence~\cite{fordbunch}.
As long as the index $\nu$ is slightly different from $3/2$, this
slight departure from scale invariance introduces a natural infrared
regularization.
To see
this, we split the integral as in Ref.~\cite{boyan1} as
\bea\label{intsplit}
\int^{\Lambda_p}_0 \frac{dz}{z} \; z^3 \; |H^{(1)}_\nu(z)|^2&=&\int^{\mu_p}_0 \frac{dz}{z} \; z^3 \, \left|H^{(1)}_\nu(z)\right|^2
\nonumber\\
&&+ \int^{\Lambda_p}_{\mu_p} \frac{dz}{z} \; z^3 \,
\left|H^{(1)}_\nu(z)\right|^2 ,\nonumber\\
\eea
where  $\mu_p$ serves
as  the cutoff for the  integral in the infrared regime to give the dominant contribution, whereas the second integral can be
absorbed by the counterterm by defining the renormalized  $\langle \varphi^2 \rangle_R $ in the renormalization scheme we choose.
In the limit of  $\Delta \rightarrow 0$ for $\nu=3/2-\Delta$, we can obtain the leading-order contributions  from
 the pole~\cite{boyan1}, by
using the small argument limit of the Hankel functions. This yields
\be\label{QC}
\frac12 \langle[\varphi(\vx,t)]^2\rangle_R = \left(\frac{H_0}{4 \, \pi}\right)^2
\left[ \frac1{\Delta}
 + 2 \, \gamma - 4 + \mathcal{O}(\Delta) \right]\,,
\ee
\noindent
where $\gamma$ is the Euler-Mascheroni constant.  While the
UV divergences are
regularization/renormalization scheme dependent, the pole in $\Delta$ arises from
the  infrared behavior and is independent of the regularization/renormalization
scheme. Later, the Hubble flow parameters will be parametrized based upon the work in Ref.~\cite{hu}, which in some regime of interest here gives
$\Delta$ small, so as to have large enhancement from $\langle \varphi^2 \rangle_R$ in the power spectrum of primordial perturbations during inflation.
With the same renormalization prescription to $\langle \varphi^2 \rangle_R$, we can define the renormalized time derivatives  and gradient terms
with their leading-order results in the limit of the small $\Delta$ obtained as
\bea\label{kinterm_R}
&&\left\langle
 \dot{\varphi}^2 \right\rangle_R = \frac{H^4_0}{8
\,\pi} \; \int^{\Lambda_p}_{0} \frac{dz}{z} \;  z^2 \;
\left|\frac{d}{dz}\left[z^{\frac{3}{2}} H^{(1)}_{\nu}(z) \right]
\right|^2 \nonumber\\
&&\quad=\frac{H^4_0}{16 \,\pi^2}\Big[ 2 \, \Delta \, \big( 1+ \, \Delta \, ( \, \ln \mu_p -\ln4-4 + 2 \, \gamma +2 \ln 4 )\nonumber\\
&&\quad\quad+\,\mathcal{O}(\Delta^2) \big)\Big] ,\nonumber\\
\label{grad} &&\left\langle\left(\frac{\nabla
\varphi}{a(t)}\right)^2 \right \rangle_R = \frac{H^4_0}{8 \,\pi} \;
\int^{\mu_p}_{0} \frac{dz}{z} \; z^{5} \;  \left|
H^{(1)}_{\nu}(z)  \right|^2
\nonumber\\
&&\quad=\frac{H^4_0}{8 \,\pi^2}\, \Big[ \mu_p^2 \, \big(1 + \Delta \, ( 2 \, \ln \mu_p -5+ 2 \, \gamma +\, \ln 4)+ \mathcal{O}(\Delta^2) \big)\Big].\nonumber\\
\eea
They do not have infrared divergences for $\nu = 3/2 $ due to
the two additional powers of the loop momentum in the integral.

{The backreaction effects from the one-loop contribution in Eq.~(\ref{phi_loop}) after the proper renormalization \cite{boyan1} can be written in terms of the renormalized $\langle \varphi^2 \rangle_R$ as
\bea \label{phi_loop_r}
&&\ddot{\Phi}_0+3\,H \,\dot{\Phi}_0 \bigg[1- \frac{H_0^2}{ 12 M^2_{Pl} \epsilon_1}\bigg(\frac{3}{2} \epsilon_2 \epsilon_3+\epsilon_2^2 \epsilon_3\bigg) \frac{\langle \varphi^2 \rangle_R }{H_0^2} \bigg] \nonumber\\
&&\quad\quad \quad\quad\quad\quad \quad\quad\quad\quad \quad\quad +V'(\Phi_0)=0 \, ,
\eea
allowing us to define the effective $\epsilon_{1 (\text{1-loop})}$
to be
\be \label{epsilon1_q}
{\epsilon_{1( \text{1-loop})}}={\epsilon_{1}} \bigg[1- \frac{H_0^2}{ 6\, M^2_{Pl} \, \epsilon_1}\bigg(\frac{3}{2} \epsilon_2 \epsilon_3+\epsilon_2^2 \epsilon_3\bigg)  \frac{\langle \varphi^2 \rangle_R }{H_0^2} \bigg]
\ee
with one-loop contributions given by $V'''(\Phi_0)$ that will be taken into account while computing the power spectrum of primordial perturbations that also include the one-loop effects of the same order of magnitude. }

\section{Power Spectrum of Primordial Perturbations} \label{sec3}
The power spectrum of primordial perturbations is described by
 the
density perturbations with  this gauge-invariant
quantity
\begin{equation}
\zeta_{ k} = \left .\frac{\delta \rho_k }{\rho_{\phi}+p_{\phi}} \right |_{\vert k \vert \leq aH}
\,  \label{masspert}
\end{equation}
evaluated in a spatially flat gauge \cite{Lee}.
The  density fluctuations $\delta \rho$ originated from the
field fluctuations $\varphi$  can be derived with the energy density
in the Friedmann equation (\ref{frieman_q}) subtracting  its classical counterpart (\ref{rho_p}), which is given by
\begin{eqnarray}
{\delta \rho} &=& \dot \Phi_0 \dot \varphi + V'(\Phi_0) \varphi +\frac{1}{2} \, {\dot{\varphi}}^2+\frac{1}{2}
 \frac{(\partial_i \varphi \partial_i \varphi)}{a^2} \nonumber\\
 &&+\frac{V''(\Phi_0)}{2 !} \varphi^2 +\frac{V'''(\Phi_0)}{3!} \varphi^3 + \textmd{higher orders in}\,\varphi .\, \nonumber\\
\label{deltadensity}
\end{eqnarray}
{In addition, summing up the energy
density and the pressure, $\rho+p$, gives
\begin{eqnarray}
 {\rho_{\phi}+p_{\phi}}=\dot{\Phi}_0^2+
 \langle \dot\varphi^2 \rangle+ \frac{1}{ a^2} \langle \partial_i \varphi \partial_i \varphi \rangle .\label{rho+p}
\end{eqnarray}
Apparently, the first term $\dot{\Phi}_0^2$ in Eq.~(\ref{rho+p})
comes from the background inflaton field. The other terms,
however, are the contributions from the quantum corrections.}

Later, all the derivatives of $V(\Phi_0)$ as a function of $\Phi_0$ can be expressed in terms of the Hubble flow parameters via Eq.~(\ref{eps_V}).
Finally,  the power spectrum  can be computed from the variable $\zeta$ as
\begin{align} \label{PS}
\Delta^2_{\zeta}(k) =\frac{k^3}{ 2\pi} \langle \zeta_{-k} \zeta_k \rangle= (\zeta^2 )_k \,.
\end{align}
In the USR inflation with a very flat inflaton potential
such that $\dot \Phi_0 \gg V'(\Phi_0)/3H_0$,  ignoring the quantum corrections can approximate the energy density fluctuations  and the sum of the energy density and the pressure  as $\delta \rho \approx \dot \Phi_0 \, \dot\varphi$ and $\rho_{\phi}+p_{\phi} \approx \dot\Phi_0^2$, respectively.
Then, the power
spectrum (\ref{PS}) becomes
\be\label{PSUSR}
\Delta^2_{\zeta, USR}(k) =\frac{(\dot\varphi^2)_k}{\dot \Phi_0^2} \,.
\ee
Also, in the limit of $\nu=3/2$,
the mode functions are given by the Hankel function with order $\nu=3/2$ as
\be
S_{\nu=3/2}(k,\eta)= \frac{H_0}{\sqrt{2 \pi}} \frac{i}{k^{3/2}} \, (-i-k \vert \eta \vert)  \; e^{-ik \vert\eta \vert}\, .
\ee
Substituting the solutions of the mode function into Eq.~(\ref{PSUSR}) and using Eq.~(\ref{V_app}) to replace $\dot \Phi_0$ by the Hubble flow parameter $\epsilon_1$, it is straightforward to achieve the standard result of the power spectrum in the USR approximation, given by
\be\label{PSUSR_S}
\Delta^2_{\zeta, USR} =\frac{H_0^2}{ 8\pi^2 \, M_{Pl}^2 \, \epsilon_1} \,.
\ee

Since the order of the Hankel function $\nu$ can be slightly deviated from $\nu=3/2$, here we provide a more involved expression that takes the deviation into account.
To do so, we  approximate ${\delta \rho} \approx \dot \Phi_0 \dot \varphi + V'(\Phi_0) \varphi $ in Eq.~(\ref{deltadensity}), and also $\rho_{\phi}+p_{\phi} \approx \dot\Phi_0^2$ as above.
The Hankel function of order $\nu$ in Eq.~(\ref{BDS}) is applied   to compute the power spectrum (\ref{PS}) instead, which is explicitly given by
\bea \label{PSUSR_BS}
&&\Delta^2_{\zeta, \nu\, USR}\nonumber\\
&&\,\,= \frac{H_0^2}{ 8\pi^2 \, M_{Pl}^2 \, \epsilon_1} \Bigg\{ \left| (-k\eta) \frac{d}{d(-k \eta)} \bigg[  (-k \eta)^{3/2} H_{\nu}^{(1)} (-k \eta)\bigg] \right|^2 \nonumber\\
&&\quad+ 3 \, \left(1+\frac{\epsilon_2}{6}\right) \, \left[ (-k\eta) \frac{d}{d(-k \eta)} \left|  (-k \eta)^{3/2} H_{\nu}^{(1)} (-k \eta) \right|^2 \right] \nonumber\\
&&\quad +9 \, \left(1+\frac{\epsilon_2}{6}\right)^2 \, (-k \eta)^3 \left| H_{\nu}^{(1)} (-k \eta) \right|^2 \Bigg\}\; ,
\eea
where $\dot\Phi_0$ and the derivatives of $V(\Phi_0)$  have been replaced by the Hubble flow parameters via Eq.~(\ref{V_app}). In the next section, the values of the Hubble flow parameters will be chosen as exemplified in a toy model in Ref.~\cite{hu}. We can then use the above expression (\ref{PSUSR_BS})  with a best chosen value of the horizon crossing time, $ \vert k\eta \vert \lesssim 1$, to compare with the power spectrum obtained by numerically solving the Mukhanov-Sasaki (MS) equation for the curvature mode functions also with the boundary conditions defined by the Bunch-Davies vacuum at $\vert k \eta \vert \ll 1$.

In this article, we will also explore the  one-loop effects from quantum fluctuations of the inflaton field itself to the power spectrum $\Delta^2_{\zeta}$ during the USR inflation.
{Nevertheless, the one-loop contributions can be obtained by Wick contraction to simply factorize $\varphi^3$ as
\be
\varphi^3 \rightarrow 3 \langle \varphi^2 \rangle \, \varphi \,,
\ee
where $\langle \varphi^2 \rangle$ is computed from the  free-field equations in (\ref{heiseqn}) with the result in (\ref{phi2}) \cite{boyan3}. }
Then, the Fourier transform of ${\delta \rho}$ in terms of the Fourier transform of $\varphi$ can be given by
\bea
{\delta \rho}_k &=& \dot \Phi_0 \dot \varphi_k +\bigg( V'(\Phi_0)+\frac{1}{2} V'''(\Phi_0) \langle \varphi^2 \rangle \bigg) \varphi_k  \, \nonumber\\
&&+ \frac{1}{2} \int \frac{d^3 k_1}{( 2\pi)^3} \, \bigg[  \dot\varphi_{k_1} \dot\varphi_{k_1 -k}+ \frac{\vec k_1 \cdot (\vec k-\vec k_1)}{2 a^2} \varphi_{k_1} \varphi_{k-k_1} \nonumber\\
&&+\frac{V''(\Phi_0)}{2\, !}  \varphi_{k_1} \varphi_{k-k_1}\bigg] \, .
\label{deltadensity_k}
\eea
In particular, using Wick contraction again,   the contribution of the $V''(\Phi_0)$ term to the power spectrum  gives
\
\begin{widetext}
 \bea
&& \int \frac{d^3 k_1}{( 2\pi)^3} \int \frac{d^3 k_2}{( 2\pi)^3} \langle  \varphi_{k_1} \varphi_{k-k_1}  \varphi_{k_2} \varphi_{-k-k_2} \rangle  =2  \int \frac{d^3 k_1}{( 2\pi)^3} \langle \varphi_{k_1} \varphi_{-k_1} \rangle \langle \varphi_{k-k_1} \varphi_{-k+k_1} \rangle  \simeq 2 \, \langle \varphi^2 \rangle  \, \langle \varphi_{k} \varphi_{-k} \rangle \, .
\eea
\end{widetext}
The momentum integral is found to be dominated in the regime of small $\vert \vec k_1 \vert$   in the case of small $\Delta$ as in Eq.~(\ref{intsplit}), and thus it can be further approximated
by involving $\langle \varphi^2 \rangle$ with large infrared effects encoded in the renormalized $\langle \varphi^2 \rangle_R$ shown in (\ref{QC}). However, due to  the lack of  infrared enhancement from  the time derivative and gradient terms as seen in Eq.~(\ref{kinterm_R}), they  will be ignored as compared with $\langle \varphi^2 \rangle_R$.

 With the definition of
\bea \label{rho2}
(\delta \rho^2 )_k && = \frac{k^3}{2 \pi} \langle \delta \rho_{-k} \, \delta \rho_k \rangle \, , \nonumber\\
( \varphi^2 )_k && = \frac{k^3}{2 \pi} \langle \varphi_{-k} \, \varphi_k \rangle \, ,
\eea
involving the one-loop quantum corrections to $(\delta \rho^2 )_k$ given by (\ref{deltadensity_k}) leads to
\begin{widetext}
\bea \label{rho_q}
(\delta \rho^2 )_k  =  \dot\Phi_0^2 \, (\dot \varphi^2)_k + \dot \Phi_0 \bigg( V'(\Phi_0) + \frac{V'''(\Phi_0)}{2} \langle \varphi^2 \rangle_R \bigg) \frac{d}{d t} (\varphi^2)_k + \bigg( V'^2 (\Phi_0) + \big(V'(\Phi_0) V'''(\Phi_0)+\frac{1}{2} V''^{\, 2} (\Phi_0) \big) \langle \varphi^2 \rangle_R \bigg) (\varphi^2)_k\, ,\nonumber\\
\eea
\end{widetext}
where the renormalized $\langle \varphi^2 \rangle_R$ is included only.
{Thus, the one-loop power spectrum  with the effects from the infrared enhanced $\langle \varphi^2 \rangle_R$ can be obtained from Eqs.~(\ref{rho_q}) and (\ref{rho+p}) with $
 \rho_{\phi}+p_{\phi} \approx \dot{\Phi}_0^2$ and
also together with the one-loop modified $\epsilon_1$ in Eq.~(\ref{epsilon1_q}) due to the backreaction  to the equation of motion for the inflaton field $\Phi_0$, giving}
\begin{widetext}
\bea \label{PSUSR_loop}
 \Delta^2_{\zeta, {\rm 1-loop}} &\simeq &\frac{H_0^2}{ 8\pi^2 \, M_{Pl}^2 \, \epsilon_1 \left[ 1 - \frac{H_0^2}{ 6\, M^2_{Pl} \, \epsilon_1}\Big(\frac{3}{2} \epsilon_2 \epsilon_3+\frac{1}{2} \epsilon_2^2 \epsilon_3\Big)  \frac{\langle \varphi^2 \rangle_R }{H_0^2} \right]}\Bigg\{  \left\vert  (-k\eta) \frac{d}{d(-k \eta)}  \bigg[ (-k \eta)^{3/2} H_{\nu}^{(1)} (-k \eta)\bigg] \right\vert^2\nonumber\\
  && +  \bigg(3\, \big(1+\frac{\epsilon_2}{6}\big) +\frac{H_0^2}{8 M_{Pl}^2 \epsilon_1} ( 3\epsilon_2 \epsilon_3+\epsilon_2^2 \epsilon_3)  \frac{\left\langle
 {\varphi}^2 \right\rangle_R}{H_0^2}
 \bigg) \, \left[ (-k\eta) \frac{d}{d(-k \eta)} \left\vert  (-k \eta)^{3/2} H_{\nu}^{(1)} (-k \eta) \right\vert^2 \right] \nonumber\\
&&+\bigg( 9 \, \big(1+\frac{\epsilon_2}{6}\big)^2+ \frac{H_0^2}{ M_{Pl}^2 \epsilon_1} \bigg[\frac{3}{4} \big(1+\frac{\epsilon_2}{6} \big) \big( 3\, \epsilon_2 \epsilon_3 +\epsilon_2^2 \epsilon_3\big) +\frac{1}{8} \big( \epsilon_2 \epsilon_3+\frac{\epsilon_2^2}{2}+ 3 \epsilon_2 \big)^2\bigg] \frac{ \left\langle
 {\varphi}^2 \right\rangle_R}{H_0^2} \bigg)  \bigg[ (-k \eta)^3 \vert H_{\nu}^{(1)} (-k \eta) \vert^2 \bigg]\Bigg\}.\nonumber\\
\eea
\end{widetext}
Assuming that $\nu =3/2 -\Delta$ and $0< \Delta <1$ during the whole course of the inflation, the renormalized $\langle \varphi^2 \rangle_R$ ~(\ref{QC}) in the small $\Delta$ approximation can be applied.
Also, notice that the quantum corrections are of order $H_0^2/( M_{Pl}^2 \epsilon_1)$. With the fine-tuning of the Hubble flow parameters, $H_0^2/(M_{Pl}^2 \epsilon_1)$ can be of order $H_0^2/(M_{Pl}^2 \epsilon_1) \sim 10^{-9}$, which is consistent with the Planck observations during the early stage of inflation, and then blows up to $H_0^2/(M_{Pl}^2 \epsilon_1) \sim 10^{-2}$
in the late-time inflation, which is large enough to sufficiently seed PBHs as well as make
the quantum corrections potentially significant.
\section{Numerical examples}\label{sec4}
To perform the numerical study, we adopt the SR step model proposed in Ref.~\cite{hu}. In the reference, the choice of the Hubble flow parameters is chosen in an effective theory approach given by
\be \label{stepmod}
\ln \epsilon_1 (N) = C_1 + C_2 N -C_3 \Big[ 1+ \tanh \Big(\frac{N-N_s}{d} \Big) \Big] \, .
\ee
 The parameters $C_1$ and $C_2$ are determined to be consistent with the scalar tilt ($n_s\approx 0.968$) and the tensor-to-scalar ratio ($r< 0.10$) in the early stage of inflation ($N < 7$). Then, $\ln \epsilon (N)$  undergoes a transition at $N=N_s$ from SR to USR  with its change, namely, $\delta \ln \epsilon_1 \sim -2 C_3$ within the width of $d$ $e$-folds.
 The inflation ends, say, at $N=60$. In Fig.~\ref{fig1}, the set of the parameters in the step model in Eq.~(\ref{stepmod}) is given by Ref.~\cite{hu}  as
$(C_1,C_2,C_3,N_s,d)=(-5.07,0.0914,8.7,40,10)$, which
allows us to compute the power spectrum. As expected, the power spectrum starts from $10^{-9}$ during the small $N$ and increases to $10^{-2}$ in the large $N$.  Also,  the more involved approximate expression of the power spectrum $\Delta^2_{\zeta,\nu USR}$ in  Eq.~(\ref{PSUSR_BS}) evaluated at the proper value, say, { $\vert k \eta \vert =0.15 <1$}, can provide a better approximation to the power spectrum  $\Delta^2_{\zeta MS}$ obtained from the exact numerical solutions of the MS equation at $\vert k \eta \vert \ll1$ in Ref.~\cite{hu} than the standard power spectrum $\Delta^2_{\zeta,USR}$  in Eq.~(\ref{PSUSR_S}).
Similar comparisons are also made with the set of the parameters $(C_1,C_2,C_3,N_s,d)=(-5.07,0.0914,8.7,40,4)$ with a more rapid transition that certainly violates the SR approximation.  Both approximate expressions $\Delta^2_{\zeta,USR}$ in Eq.~(\ref{PSUSR_S}) and  $\Delta^2_{\zeta,\nu USR}$ in Eq.~(\ref{PSUSR_BS}) show relatively large discrepancies with $\Delta^2_{\zeta,MS}$ for such a narrow width $d$, as shown in Fig.~\ref{fig2}.
\begin{figure}[t]
\begin{center}
\includegraphics[width=1.\linewidth]{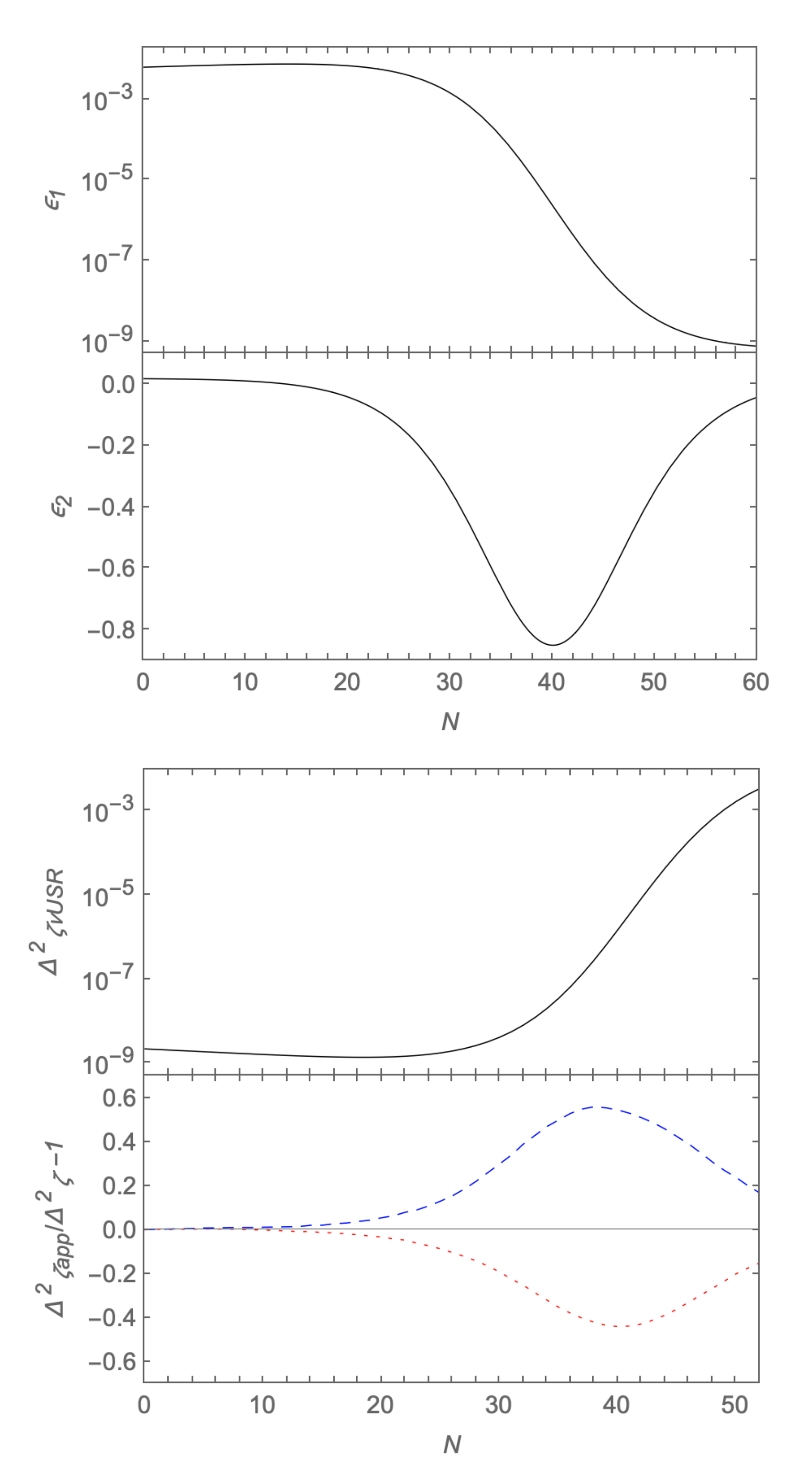}
\caption{The parameter set in the step model in Eq.~(\ref{stepmod}) is chosen by following Ref.~\cite{hu}  as
$(C_1,C_2,C_3,N_s,d)=(-5.07,0.0914,8.7,40,10)$. The evolution of the Hubble flow parameters $\epsilon_1$ and $\epsilon_2$ is shown in the upper panel. In the bottom panel, the power spectrum $\Delta^2_{\zeta,\nu,USR}$ is plotted according to Eq.~(\ref{PSUSR_BS}) when $\vert k \eta \vert=0.2$. The deviation of the approximate result $\Delta^2_{\zeta,\nu,USR}$ from the power spectrum $\Delta^2_{\zeta, MS} $ obtained by numerically solving the MS equation in Ref.~\cite{hu} is plotted as the red dotted line, whereas the deviation from $\Delta^2_{\zeta, MS} $ for the standard expression $\Delta^2_{\zeta,USR}~$  (\ref{PSUSR_S}) is also shown with the blue dashed line for a comparison.  }
\label{fig1}
\end{center}
\end{figure}

\begin{figure}[t]
\begin{center}
\includegraphics[width=1\linewidth, trim=0 0 0 25]{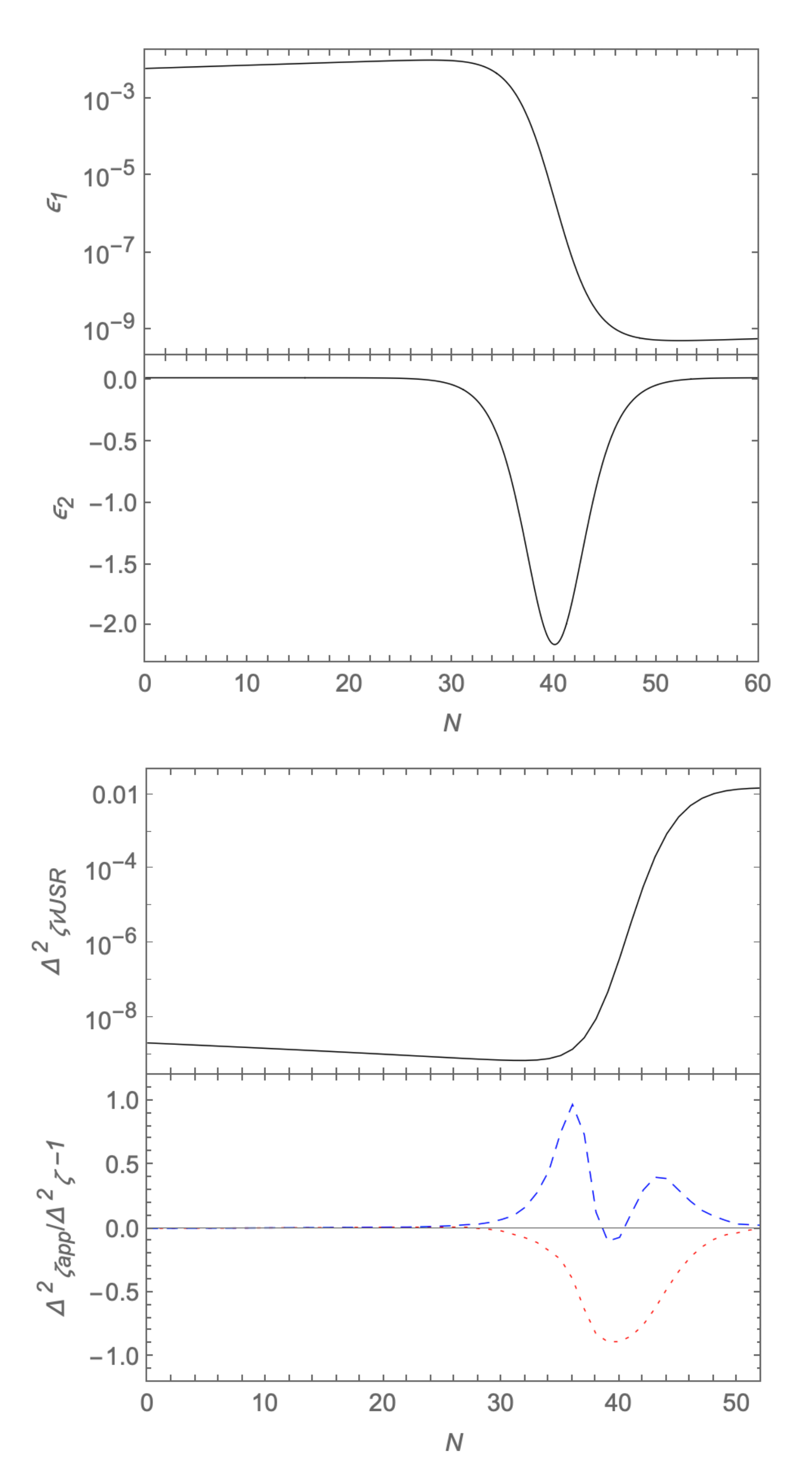}
\caption{Same as in Fig.~\ref{fig1}, but with the parameter set, $(C_1,C_2,C_3,N_s,d)=(-5.07,0.0914,8.7,40,4)$. }
\label{fig2}
\end{center}
\end{figure}
Next, we will include the one-loop effects to the power spectrum $\Delta^2_{\zeta, {\rm 1-loop}}$ in Eq.~(\ref{PSUSR_loop}).
Here, we choose the parameters so that during the whole course of inflation the order of the Hankel function $\nu=3/2-\Delta$ almost remains $\nu\sim 3/2$, namely, $0< \Delta <1$.  The large enhancement will be seen to boost the power spectrum.
\begin{figure}[t]
\begin{center}
\includegraphics[width=1.\linewidth]{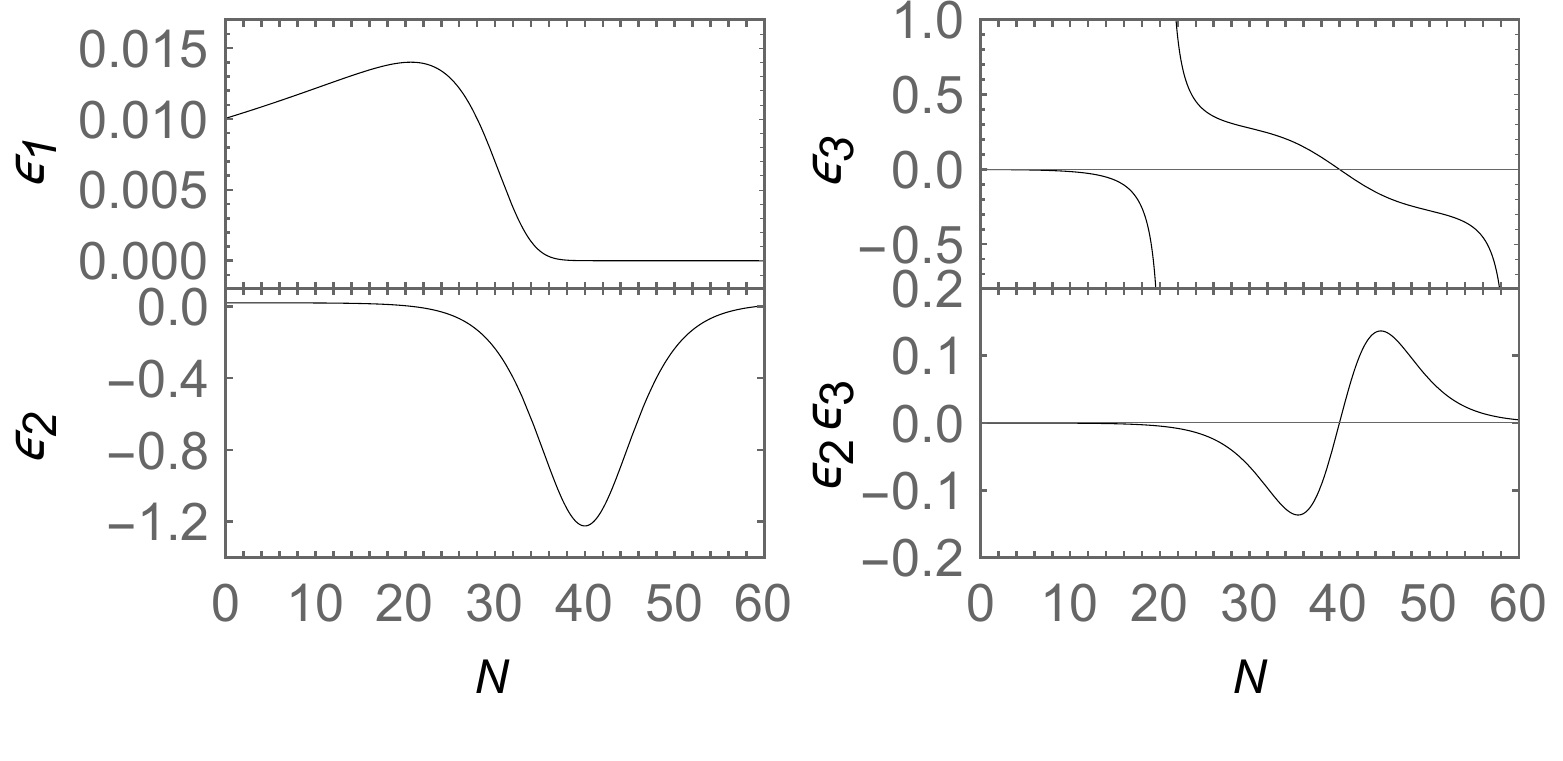}
\caption{Evolution of the Hubble parameters ($\epsilon_1,\epsilon_2,\epsilon_3,\textcolor{blue}{\epsilon_3 \epsilon_2}$) for the set of $(C_1,C_2,C_3,N_s,d)=(-4.6,0.0914,8.7,40,7)$. }
\label{fig3}
\end{center}
\end{figure}
\begin{figure}[t]
\begin{center}
\includegraphics[width=1.\linewidth]{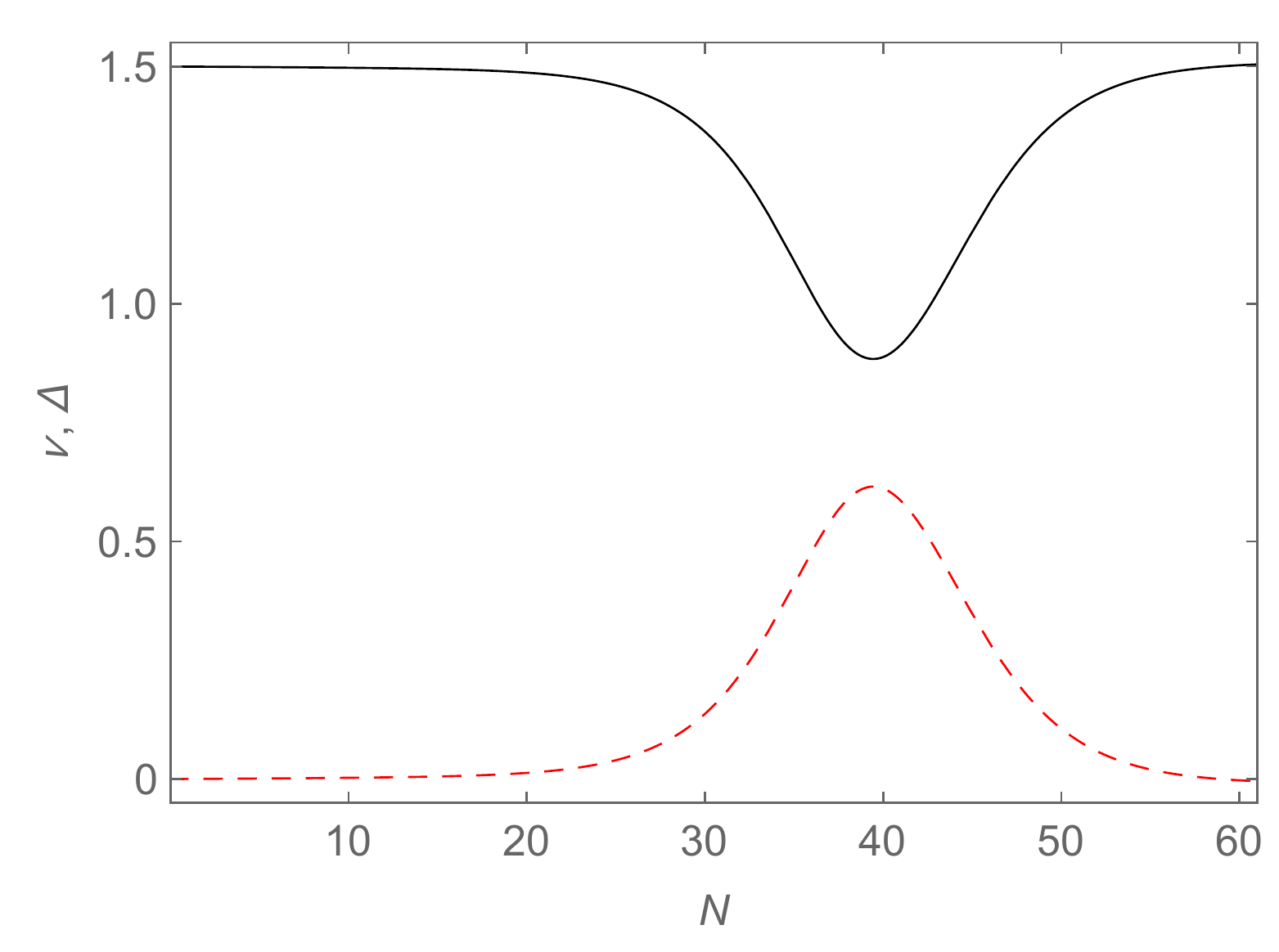}
\caption{With the same parameters in Fig.~\ref{fig3}, we show that  the evolution of the order of the Hankel function $\nu$ in Eq.~(\ref{nu}) (black solid line) and the value $\Delta$ (red dashed line) defined as $\nu=3/2-\Delta$ as  a function of $N$. }
\label{fig4}
\end{center}
\end{figure}
To do so, we choose $(C_1,C_2,C_3,N_s,d)=(-4.6,0.0914,8.7,40,7)$,  where $C_1$ is slightly changed, but the choice of the value still satisfies the Planck constraints. Also, the width $d$ is picked for the sake of clear illustration to be $d=7$  in the case of relatively wide width. In Fig.~\ref{fig3},  the evolution of the Hubble parameters with the above choice of the parameters is plotted.  $\epsilon_1$ is small in the small $N$ and drops to an extremely small value at the transition $N=N_s=40$,  entering the USR inflation. As a result,  $\epsilon_2$, as it measures the change of the $\epsilon_1$ in Eq.~(\ref{HubFlow}), goes to a negative value with a large absolute value of order $\vert \epsilon_2 \vert \sim \mathcal{O} (1)$ to have the $\dot \Phi$ term dominated although it is not an extreme case with $\epsilon_2\rightarrow -6$ for a very flat potential.
When $\epsilon_2$ goes from positive to negative values, in particular,  crossing zero,
$\epsilon_3$ then becomes large, and  $\vert \epsilon_2 \epsilon_3 \vert > \epsilon_1$. Notice that, with this choice of the parameters during the USR regime, the arguments in Ref.~\cite{CRU}, stating that $\epsilon_3 \simeq 2 \epsilon_1$ with both being small, seem not to hold. Whether or not the conclusions drawn in Ref.~\cite{CRU} are still true deserves further study. In Fig.~\ref{fig4}, we  show the order of the Hankel function $\nu$ and the value $\Delta$  as a function of $N$, lying within $0< \Delta <1$, expected to induce the significant enhancement from the one-loop contributions. Also, although the values of $\epsilon_3$ have dramatic changes in some $N$,  $\epsilon_2 \epsilon_3$ changes smoothly as  $N$ increases, consistent with the adiabatic  approximation since the dependence of $\epsilon_3$ in the quantities we compute here  always appears in a way of $\epsilon_2 \epsilon_3$. Finally, using the result in Eq.~(\ref{PSUSR_loop}) and the power spectrum in Eq.~(\ref{PSUSR_BS}) (which is plotted in Fig.~\ref{fig5}), the correction to the power spectrum from the one-loop contributions is shown in Fig.~\ref{fig6}, in which all quantum corrections are given by Eq.~(\ref{QC})  in their small $\Delta $ limit.
It is anticipated that, although $\Delta$ is small for the whole course of inflation, the infrared enhancement is significant when $\Delta $ approaches to zero at $ N\sim 60$ and also $H_0^2/(M_{Pl}^2 \epsilon_1)$ grows from $\approx 10^{-9}$ to $\approx 10^{-2}$.
In fact, the evolution of the order of the Hankel function $\nu$ in Eq.~(\ref{nu}) is mainly determined by $\epsilon_2$ and can be approximated by
{$\nu \approx (9/4+3\epsilon_2/2 +\epsilon_2^2/4)^{1/2}$.} In particular, during the transition from SR to USR inflation,  $\epsilon_1$ is driven to an extremely small value by  a negative value of  $\epsilon_2 $ of which the absolute value is relatively large.
{Then, staying in the phase of USR requires $\epsilon_1$ to maintain that small number with very little change, thus driving $\epsilon_2$ from the negative value toward zero that ends USR inflation and then enters the second  SR inflation, giving $\nu=3/2$.}
So, when the power spectrum reaches a high value, the order of the Hankel function is driven to $\nu=3/2$, and at the same time, the infrared divergence of the one-loop effects given by the minimally coupled massless scalar field in the de Sitter space-time also makes significant corrections to the power spectrum.

\begin{figure}[t]
\begin{center}
\includegraphics[width=1\linewidth, trim=0 0 0 20]{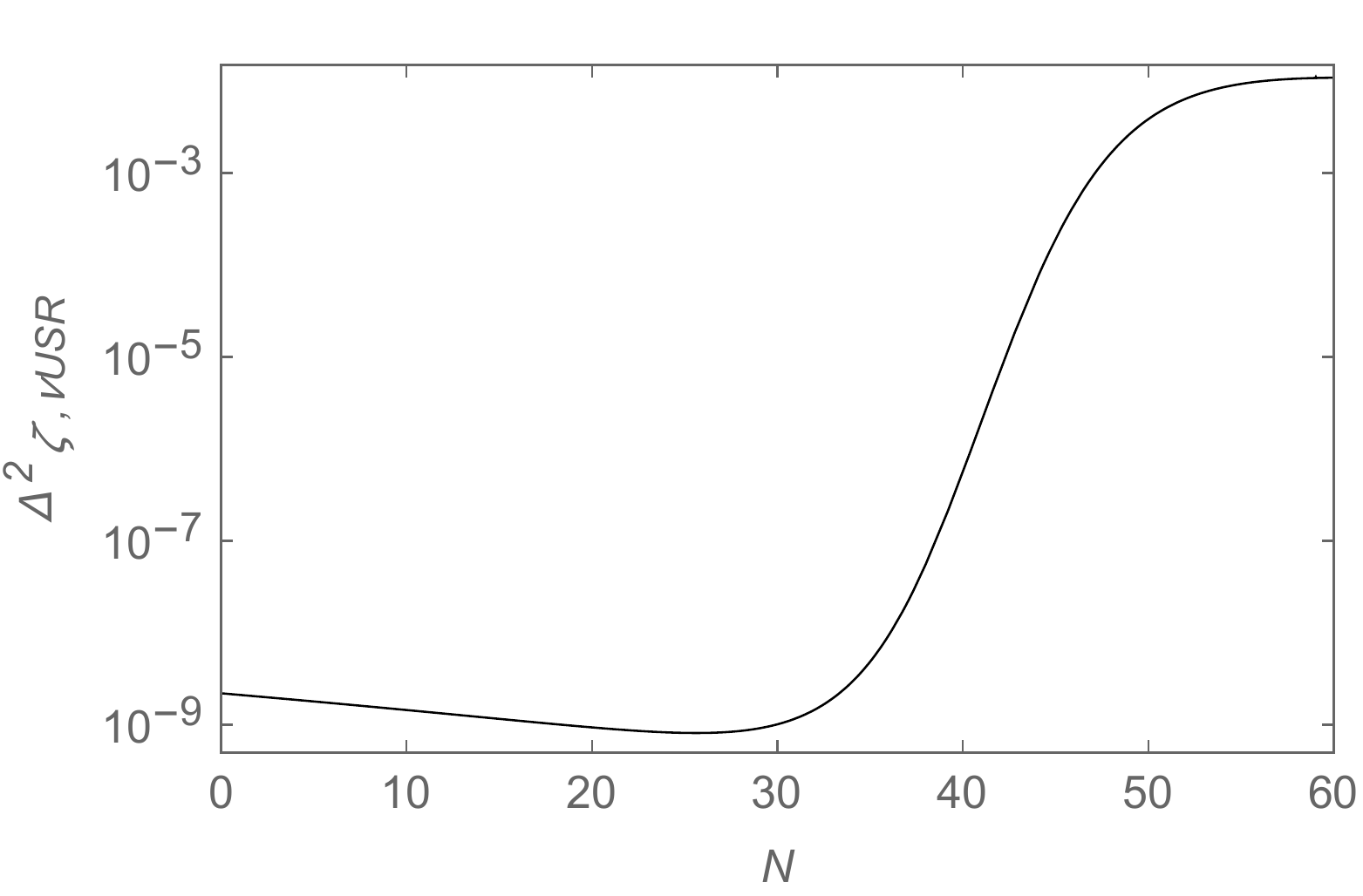}
\caption{Evolution of the power spectrum $\Delta_{\zeta, \nu USR}$~(\ref{PSUSR_BS})  as a function of $N$ for the parameter set in Fig.~\ref{fig3}.}
\label{fig5}
\end{center}
\end{figure}
\begin{figure}[b]
\begin{center}
\includegraphics[width=1\linewidth, trim=0 0 0 0]{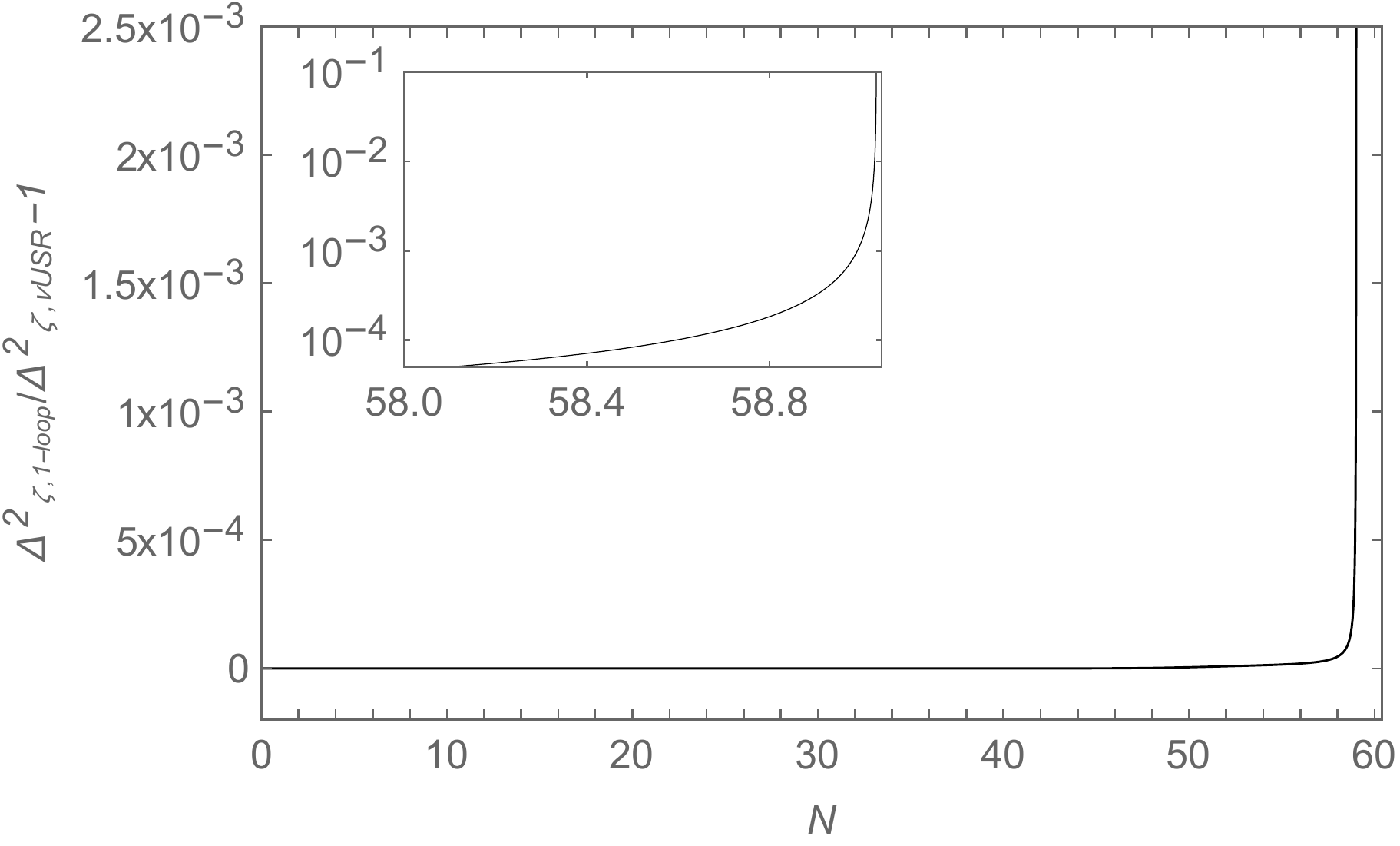}
\caption{Corrections due to the one-loop effects as a function of $N$ to the full one-loop result
$ \Delta^2_{\zeta, {\rm 1-loop}} $~(\ref{PSUSR_loop}) deviated from $\Delta_{\zeta, \nu USR}$~(\ref{PSUSR_BS}). }
\label{fig6}
\end{center}
\end{figure}
Finally, we come to study the case in which the Universe undergoes SR-USR-SR inflation. To model this scenario, we modify the above parametrization of the Hubble flow variables by adding three more parameters, $N_{\bar s}$, $\bar d$, and $C_4$, specifying  the starting $N$ when  the transition from USR back to SR occurs, the width of this transition, and the amount of the change in  $\ln \epsilon_1$, respectively.
The formula reads
\bea
\ln \epsilon_1 &=& C_1+C_2N-C_3\left[1+\tanh{\left(\frac{N-N_s}{d}\right)}\right]\nonumber\\
&&+C_4 \left[ \tan^{-1}\left(\frac{N-N_{\bar s}}{\bar d}\right)+\frac{\pi}{2}  \right] \, .
\label{sr-usr-sr}
\eea

\begin{figure}[t]
\begin{center}
\includegraphics[width=1.\linewidth]{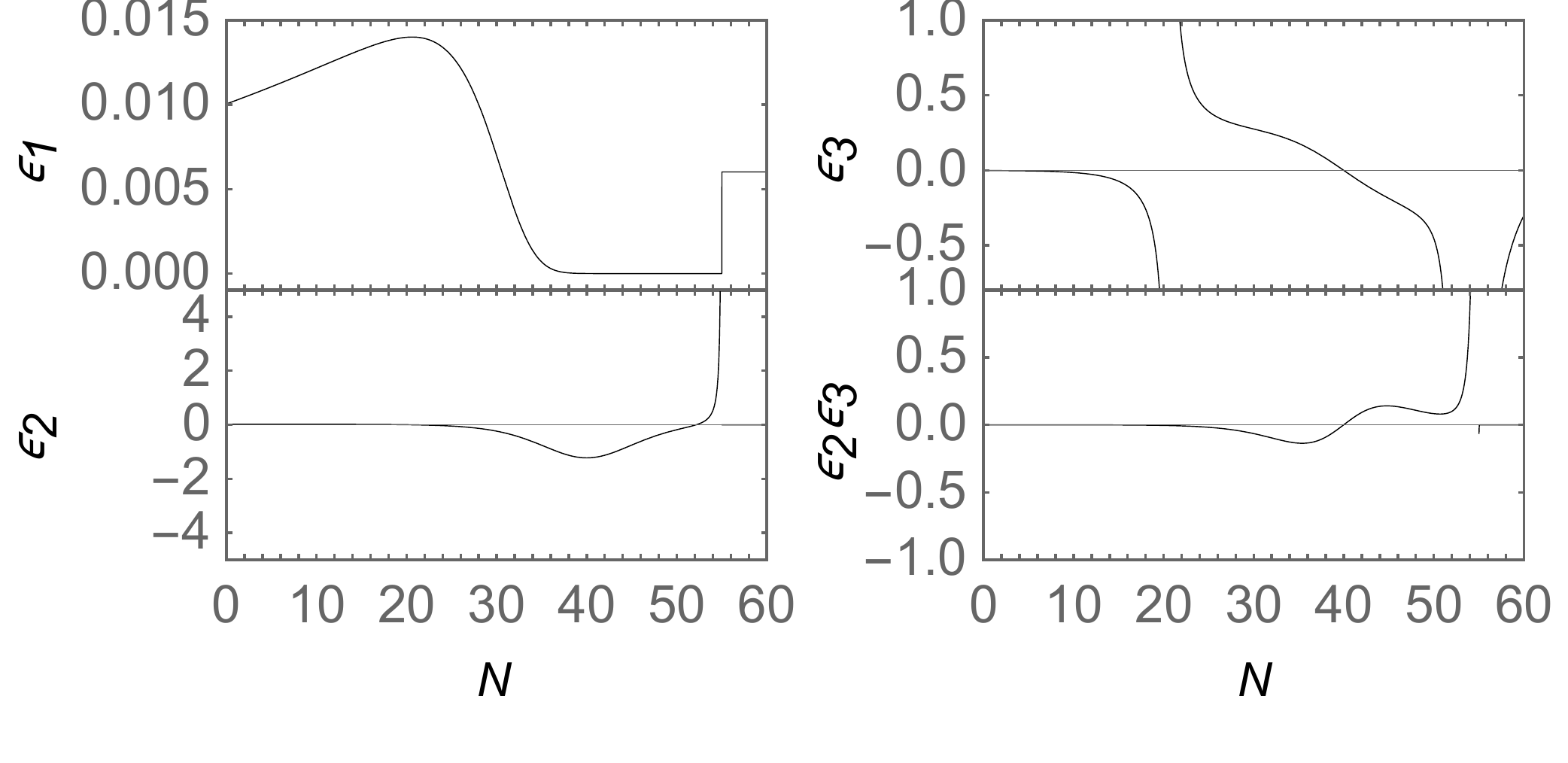}
\caption{Evolution of the Hubble parameters ($\epsilon_1,\epsilon_2,\epsilon_3,\textcolor{blue}{\epsilon_2\epsilon_3}$) as a function of $N$ for the set of $(C_1,C_2,C_3,N_s,d; C_4,\bar d, N_{\bar s})=(-4.6,0.0914,8.7,40,7;0.00192,7 \times 10^{-7}, 55)$ in Eq.~(\ref{sr-usr-sr}).}
\label{fig7}
\end{center}
\end{figure}
For a clear illustration, we fine tune the parameters for such a transition that happens at $N=60$ with a very narrow width $\bar d$, which does not intervene
with the rise to the maximum value of the power spectrum at $N_s=40$. This set of parameters is given by
 $C_1=-4.6$, $C_2=0.0194$, $C_3=8.7$, $d=7$, $N_s=40$, $C_4=0.00192$, $\bar d=7 \times 10^{-7}$, and $N_{\bar s}=55$. All parameters
 relevant to the SR to USR transition remain the same as in Fig.~\ref{fig3}.  By adding another transition from USR to SR, it is seen that the evolution of $\epsilon_1$ changes from an extremely small value back to the order of $10^{-2}$ after the transition point $N=55$.
 Then, all other Hubble flow variables end up with small values in the second SR regime, and the corresponding power spectrum settles to the small value $10^{-9}$ again, as shown in Fig.~\ref{fig9}.
 However, during the period of the transition back to SR, $\epsilon_2$ goes from a negative to a positive value and thus crosses zero seen in Fig.~\ref{fig7}.
 When $\epsilon_2$ is in the regime of $\epsilon_2 \rightarrow  0^{-}$ seen in Fig. \ref{fig7}, giving $\nu \simeq 3/2-\Delta$ with $\Delta \rightarrow 0^{+}$ where the adiabatic approximations still hold, as in Fig. \ref{fig6},  the quantum loop effects to $\langle \varphi^2 \rangle$ are enhanced due to the infrared divergence in its momentum integral in Eq.~(\ref{QC}). One of the main conclusions in this work based upon the consistent adiabatic approximations is that we find the significant one-loop corrections  around the peak of the
density power spectrum in both scenarios.

 \begin{figure}[t]
\begin{center}
\includegraphics[width=1\linewidth]{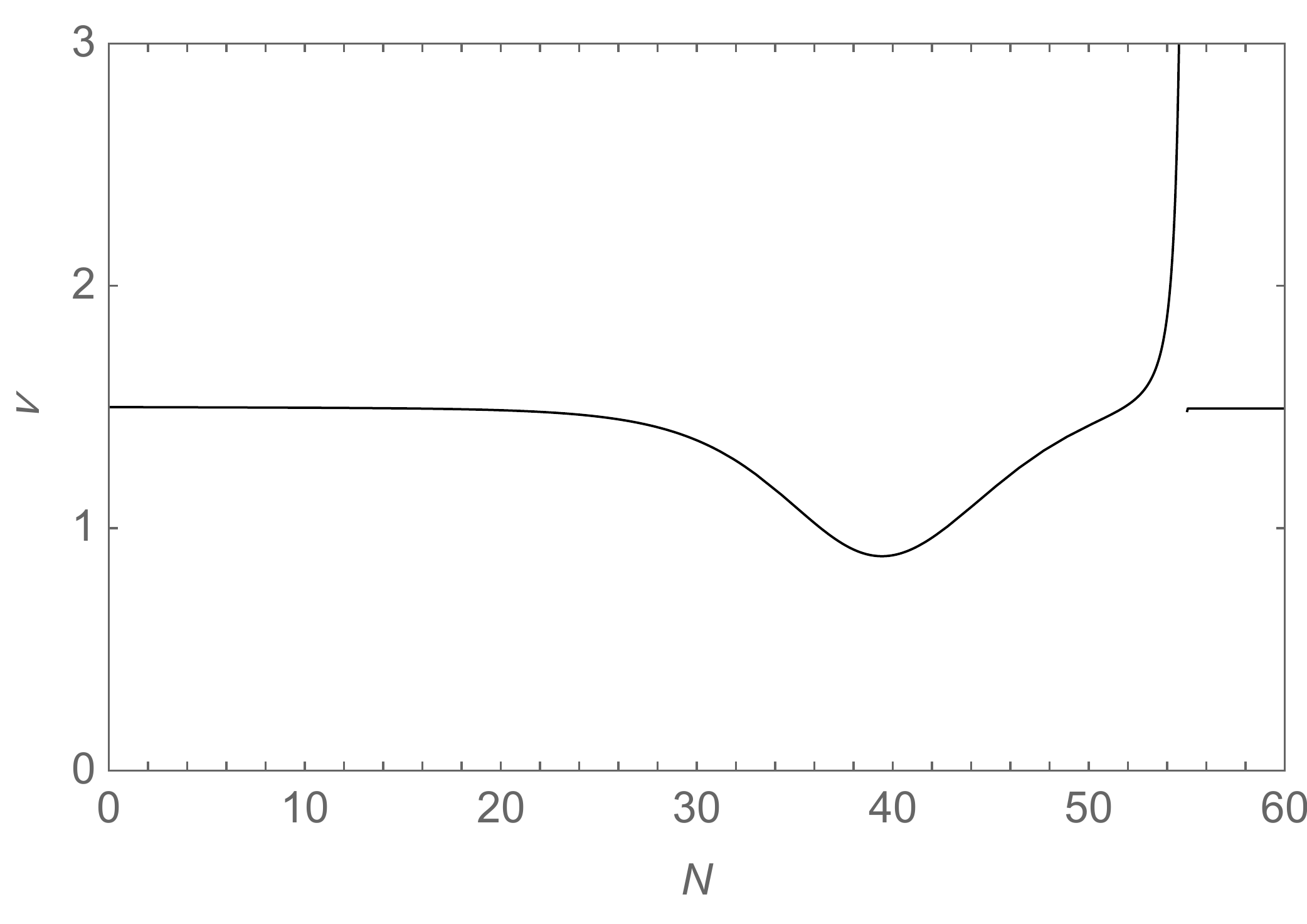}
\caption{Evolution of the order of the Hankel function $\nu$ in Eq.~(\ref{nu}) as a function of $N$ for the parameter set in Fig.~\ref{fig7}.}
\label{fig8}
\end{center}
\end{figure}

Nevertheless,  during the USR to SR transition, $\epsilon_2$ can become positive with $\nu=3/2-\Delta$, where $\Delta$ is negative seen in Fig. \ref{fig8}, although the adiabatic approximations seem to break down,
the one-loop effects of $\langle \varphi^2 \rangle$  encounter a different type of infrared divergence in Eq.~(\ref{phi2}).
We find that
 \bea  \label{phi2_IR}
\frac{4\pi^2}{H^2}   \langle \varphi^2 \rangle_R &=&\frac{\pi}{2}
\int^{\mu_p}_{\delta} \frac{dz}{z} \; z^3\,
\left|H^{(1)}_\nu(z)\right|^2 \nonumber\\
&\simeq & \int^{\mu_p}_{\delta} dz \,  z^{-1+ 2 \Delta}  + .... \,\nonumber\\
&\simeq & \frac{1}{2  \, \vert\Delta\vert  \, \, \delta^{2 \vert\Delta \vert} } +... \, ,
  \eea
where special care needs to be taken to resume all important infrared effects for having a reliable power spectrum especially around its peak value.
We will tackle this infrared issue beyond the adiabatic approximations in future work. Here, we give an intuitive way of regularizing the divergences by introducing physical cutoffs for $z$~\cite{xue}. Note that the USR to SR transition occurs in the period of $e$-folds from $N_1$ to $N_2$, which is controlled by $\bar d$ in the above parameters.
The one-loop effects of $\langle \varphi^2 \rangle$ can be roughly estimated by considering the momentum modes, which are within subhorizon modes at $e$-folds $N_1$ and leave out of the horizon at $e$-folds $N_2$. Thus, we have{
\bea \label{phi2_Delta-}
\frac{4\pi^2}{H^2}  \langle \varphi^2 \rangle_R &  \sim & \int^{\delta_2}_{\delta_1} dz \,  z^{-1- 2 \vert \Delta \vert}  + ... \nonumber\\
& =& -\frac{1}{ 2 \vert \Delta \vert }  \, \Big( \,  \delta_2^{-2 \vert\Delta \vert}-  \delta_1^{-2 \vert\Delta \vert} \, \Big) +... \, \nonumber\\
& \simeq &\ln \Bigg( \frac{\delta_2}{\delta_1}\Bigg) +\mathcal{O} (\vert \Delta \vert) \nonumber\\
& = & (N_2-N_1) +\mathcal{O} (\vert \Delta \vert) \, .
  \eea

\begin{figure}[t]
\begin{center}
\includegraphics[width=1\linewidth]{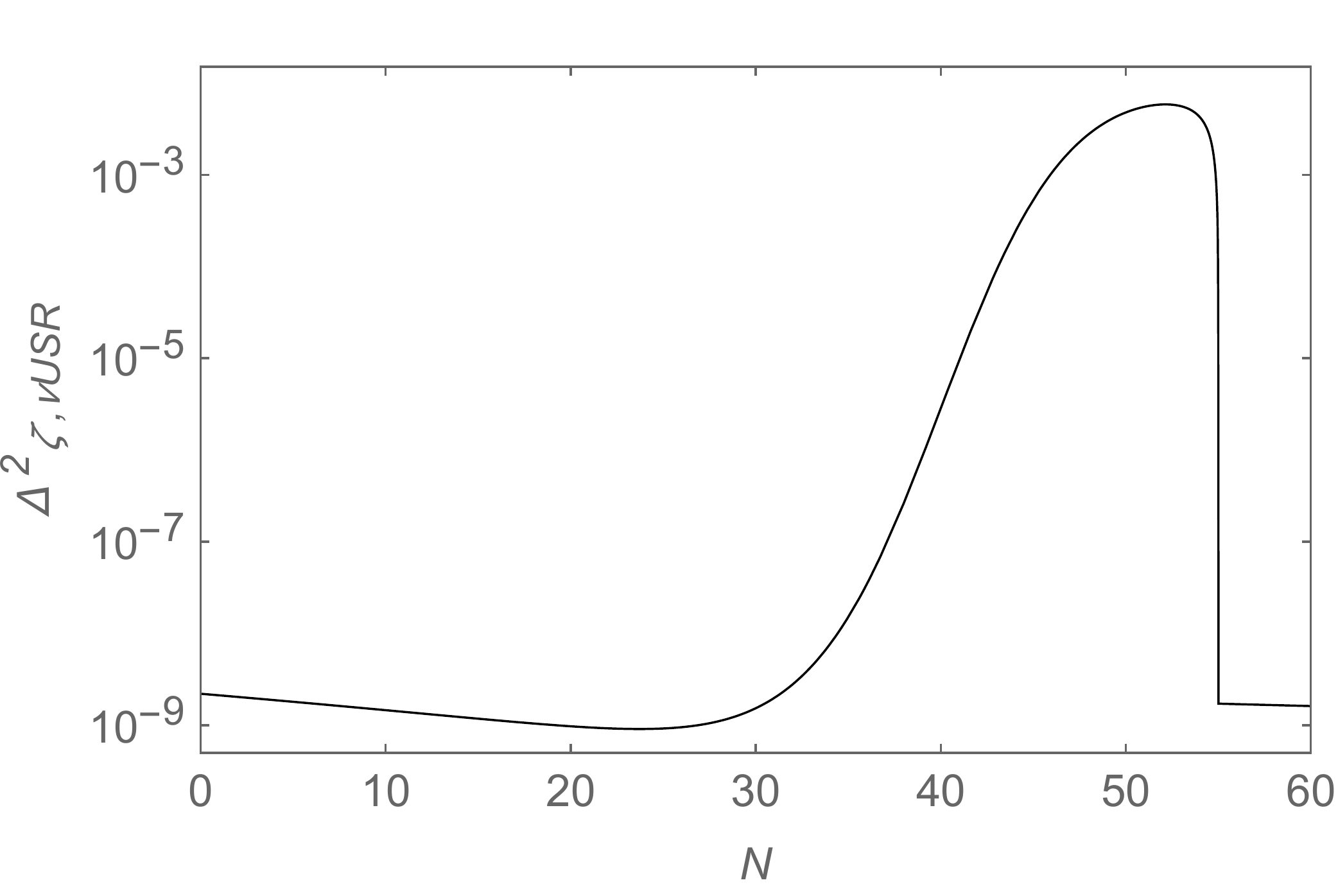}
\caption{Evolution of the power spectrum $\Delta_{\zeta, \nu USR}$~(\ref{PSUSR_BS}) as a function of $N$ for the parameter set in Fig.~\ref{fig7}.}
\label{fig9}
\end{center}
\end{figure}
In the limit of $\Delta \rightarrow 0$, the last equality of the above equation recovers the linear growth in time since $N \propto t$.
As such, in this estimation, the removal of the infrared divergence would make the one-loop effects of $\langle \varphi^2 \rangle$ ineffective, thus depending on the detailed model between USR to SR transition.
However, even for a finite $ \Delta <0$, the one-loop effects of $\langle \varphi^2 \rangle$ still suffer from infrared divergence as found  in the second line of Eq.~(\ref{phi2_IR}) as the lower limit of the momentum integral $\delta \rightarrow 0$.
According to Ref.~\cite{boyan2}, the negative value of $\Delta$ may arise from the scalar field potential with a negative mass term.
It will lead to the so-called spinodal instabilities, driving the growth of large quantum fluctuations, which needs to be incorporated by the nonperturbative method in a self-consistent manner.
Our estimate here just indicates significant one-loop corrections to the USR to SR transition that should be confirmed by a formal quantum field theoretical method.

\section{CONCLUSION}\label{sec5}
In this work, we  examine the quantum loop effects on the single-field inflationary models in a spatially flat FRW cosmological space-time in which the  general self-interacting scalar
field potential $V(\phi)$ is modeled  in terms of the Hubble flow parameters in the effective field theory approach. We
start from the metric of the perturbed  spatially flat FRW cosmological space-time in the ADM form and then  separate  the classical homogeneous background field ($\Phi_0$)  from the
quantum field fluctuations ($\varphi$).   In a spatially flat gauge, the equation of motion for the background field  in the FRW metric and the modified Friedmann equation of the scale factor with one-loop corrections are derived.
 We also derive the equation of motion for mode functions of the quantum field  in which  the solutions are  given by
the Hankel function with the order $\nu$.  The index  $\nu $ depends on the potential function of the scalar field and can be approximately expressed by the Hubble flow parameters as {$\nu^2 \simeq
 \frac{9}{4}  +( \frac{3}{2}\, \epsilon_2 +\frac{1}{4} \, \epsilon_2^2) (1+2 \epsilon_1) +\frac{1}{2} \,\epsilon_2 \epsilon_3+3 \, \epsilon_1+ \epsilon_1 \epsilon_2+6 \epsilon_1 (1+\frac{\epsilon_2}{6} )$} up to the order $\epsilon_2 \epsilon_3$ in the parameter regime where $\epsilon_1$ is  extremely small, $\epsilon_2$ is negative with $\vert \epsilon_2\vert \sim \mathcal{O} (1)$, and $\vert \epsilon_2 \epsilon_3 \vert <1$ but $\vert \epsilon_2 \epsilon_3 \vert > \epsilon_1$ during the USR inflationary epoch. { To incorporate the SR epoch in the models we propose} with $\epsilon_1 > \vert \epsilon_2\vert$ where $\epsilon_1, \vert \epsilon_2 \vert \ll1$, the linear $\epsilon_1$ terms are also included.
Later,  the one-loop contribution of $\langle \varphi^2 \rangle$ is computed for a choice of the  Bunch-Davies vacuum state.  More importantly, the renormalized $\langle \varphi^2 \rangle_R$ after subtracting the UV-divergence encounters the infrared divergence  in the case of minimally coupled massless inflaton fluctuations in de Sitter space-time as $\nu \rightarrow 3/2$ and thus is infrared enhanced for small $\Delta$ where $\nu=\frac{3}{2}-\Delta$.
In addition,  we introduce the power spectrum of primordial perturbations  described by the density perturbations in a spatially flat gauge.
 The one-loop expressions of the density perturbations as well as the energy density and pressure  of the inflaton field are obtained when the infrared enhanced $\langle \varphi^2 \rangle_R$ is considered.
Notice that the backreaction effect of $\langle \varphi^2 \rangle_R$ to the dynamics of the background inflation field also needs to be taken into account to compute the power spectrum via the one-loop modified $\epsilon_1$.
  Here, we first adopt the SR step model proposed in Ref. \cite{hu} to numerically study the SR to USR inflation. We find a huge amplification on the power spectrum of order $10^{-2}$ during the USR regime, which is large enough to potentially produce PBHs, by keeping $\epsilon_1$ ( $\epsilon_1 \sim 10^{-9}$ ) to be extremely small.
 Then, the index $\nu$ in the USR regime can be approximated by {$\nu \approx (9/4 + 3 \epsilon_2 /2 +\epsilon_2^2/4)^{1/2}$.}
 {However, staying with the small $\epsilon_1$ leads to the small $\epsilon_2$ that  ends USR inflation and later enters the second  SR inflation}, driving $\nu \rightarrow 3/2$,
 where the accompanying quantum loop effects also become significant
 as well.
 We then modify the model to consider the SR-USR-SR inflation and the peak of the power spectrum occurs in the transition of USR back to SR
   as $\epsilon_2$  goes from negative to positive
values, which then drives $\epsilon_1$ back to a relatively large value ($\epsilon_1 \sim 10^{-2}$) with the power spectrum of order $10^{-9}$ in the SR regime.   Again, when $\nu \rightarrow 3/2$,  large quantum loop effects can be seen near the peak of the power spectrum
as compared with the tree-level results. Thus, to model a successful model that undergoes either SR-USR or SR-USR-SR inflation seems inevitably
to induce large quantum loop effects. Here,
 our estimates  just indicate
significant one-loop corrections  that should be confirmed by a
formal quantum field theoretical method and, if so, further be treated in a self-consistent manner by following the works of Refs.~\cite{Lee,boyan2} in our future study.

\begin{acknowledgments}
This work was supported in part by the Ministry of Science and Technology (MOST) of Taiwan, Republic of China, under
grant No. 107-2112-M-259-003 (D.S.L.) and No. 108-2112-M-001-008 (K.W.N.).  We would like to thank D. Boyanovsky for stimulating discussions.
\end{acknowledgments}

\appendix
\section{The one-loop UV divergence}\label{appen}
In this Appendix, we summarize the UV divergence in the one-loop contributions to
the relevant quantities to the density perturbations.
The quantum correction $\langle \varphi^2 \rangle$  determined by the momentum
integral in Eq.~(\ref{phi2}) has both quadratic
and logarithmic divergences.
We introduce the UV cutoff momentum scale $\Lambda_p(\eta)$  determined by  the fixed physical cutoff
divided by  the scale of inflation in a comoving frame according to Ref.~\cite{xue}, i.e.,
$\Lambda_p(\eta)
\equiv \frac{\Lambda}{H \;  C(\eta)} \simeq -\Lambda\,  \eta
$ for small $\epsilon_1$.
For a general $\nu$, the ultraviolet divergence  of $\langle \varphi^2 \rangle$ in Eq.~(\ref{phi2}) is found to be
\be
\langle {\varphi}^2 \rangle =\frac{H_0^2}{8
\,\pi^2}\, \bigg[
{\Lambda_p^2}+ {\ln \Lambda_P} \, ( \nu^2-{1/4}) + \textmd{finite parts}\bigg] \;. \;
\ee
Similarly, the time derivatives and the gradient terms of quantum corrections also suffer from the ultraviolet divergences
 given, respectively, by
\begin{widetext}
 \bea\label{kinterm}
 \left\langle
 \dot{\varphi}^2 \right\rangle &=& \frac{H^4_0}{8
\,\pi} \; \int^{\Lambda_p}_{0} \frac{dz}{z} \;  z^2 \;
\left|\frac{d}{dz}\left[z^{\frac{3}{2}} H^{(1)}_{\nu}(z) \right]
\right|^2 \nonumber\\
&=& \frac{H^4_0}{16 \,\pi^2}\bigg[ \Lambda_p^4-\Lambda_p^2 \, (\nu^2-9/4)-\frac{\ln \Lambda_p}{2} \, (\nu^4-5\nu^2/2+9/16) +\textmd{finite parts}\bigg],\nonumber\\
\label{grad} \left\langle\left(\frac{\nabla
\varphi}{a(t)}\right)^2 \right \rangle &=& \frac{H^4_0}{8 \,\pi} \;
\int^{\Lambda_p}_{0} \frac{dz}{z} \; z^{5} \;  \left|
H^{(1)}_{\nu}(z)  \right|^2 \nonumber\\
 &=& \frac{H^4_0}{16 \,\pi^2}\bigg[ \Lambda_p^4-\Lambda_p^2 \, (\nu^2-9/4)-\frac{3 \, \ln \Lambda_p}{2} \, (\nu^4-5\nu^2/2+9/16)+\textmd{finite parts}\bigg].\nonumber\\
\eea
\end{widetext}
Here, we consider the case of the general $\nu$, and the discussions of the UV divergence above generalize those results in Ref.~\cite{boyan1}, in which the case $\nu\simeq 3/2$ is considered.
The above integrals have  the ultraviolet
divergences, which can be absorbed by the renormalization counterterms within a given renormalization scheme in the
effective field theory by following Ref.~\cite{boyan1}.

\end{document}